\documentclass[aps,reprint,floatfix,superscriptaddress,prl]{revtex4-2}
\usepackage[english]{babel}
\usepackage{epsfig}
\usepackage{amsmath}
\usepackage{enumerate}
\usepackage{bbold}
\usepackage{comment}
\usepackage{hyperref}

\usepackage{amsmath,amsfonts,bm,mathrsfs,dsfont,amssymb,xfrac,bigints,cancel,mathtools,braket}

\usepackage{soul}

\renewcommand{\vec}[1]{\mathbf{#1}}
\newcommand{\mtx}[1]{\bm#1}

\newcommand{\grad}{\vec{\nabla}}
\newcommand{\diver}{\vec{\nabla}\!\cdot}
\newcommand{\pvec}[1]{\vec{#1}\mkern2mu\vphantom{#1}}

\DeclareMathOperator{\Tr}{Tr}
\usepackage[dvipsnames]{xcolor}

\begin{document}
	
	\title{The shear viscosity of interacting graphene}
	\author{Kitinan Pongsangangan}	\affiliation{Institute for Theoretical Physics and Center for Extreme Matter and Emergent Phenomena,
		Utrecht University, Princetonplein 5, 3584 CC Utrecht, The Netherlands}
  \affiliation{Institute of Theoretical Physics, Technische Universit\"at Dresden, 01062 Dresden, Germany}
	\author{Pedro Cosme}
		\affiliation{Institute for Theoretical Physics and Center for Extreme Matter and Emergent Phenomena,
		Utrecht University, Princetonplein 5, 3584 CC Utrecht, The Netherlands}
	\affiliation{GoLP - Instituto de Plasmas e Fus\~ao Nuclear, Instituto Superior T\'ecnico, Universidade de Lisboa, 1049-001 Lisbon, Portugal}
	\author{Emanuele Di Salvo}
	\affiliation{Institute for Theoretical Physics and Center for Extreme Matter and Emergent Phenomena,
		Utrecht University, Princetonplein 5, 3584 CC Utrecht, The Netherlands}
	\author{Lars Fritz}
	\affiliation{Institute for Theoretical Physics and Center for Extreme Matter and Emergent Phenomena,
		Utrecht University, Princetonplein 5, 3584 CC Utrecht, The Netherlands}

	\begin{abstract}
	One of the hallmark properties of fluids is their shear viscosity which is, among other things, responsible for parabolic flow profiles through narrow channels. In recent years, there has been a growing number of observations of said flow profiles in electronic transport measurements in a variety of material systems, most notably in graphene. In this paper, we investigate the shear viscosity of  interacting graphene from a theoretical point of view. We study both a phenomenological as well as a microscopic model and find excellent agreement between the two. Our main finding is collective modes make a sizeable contribution to the viscosity that can equal or even outweigh the electronic contribution that is usually assumed dominant. We comment on how this finding carries over to systems beyond graphene and related Dirac materials. 
	
	\end{abstract} 
	\maketitle
Hydrodynamics describes the flow properties of classical fluids such as water, air, or plasmas. Its foundations are conserved quantities, such as mass, momentum, and energy, and their slow relaxation towards local equilibrium. This relaxation is driven by interactions between the constituents of the fluid. A direct consequence of said interactions is their shear viscosity. This manifests itself in a friction or drag between adjacent layers of the fluid in the presence of a velocity gradient \cite{LandauLifshitz1987}.

In recent years, hydrodynamics gained renewed attention across many disciplines, including condensed-matter physics \cite{Hartnoll2007,Spivak2011,Zaanen2016,Levitov2016,Lucas2018,Narozhny2019,fritz2023hydrodynamic}. It is understood as an emergent universal description of dynamical properties of strongly correlated systems in their long-wavelength and low-frequency limit \cite{Kovtun2012}. The framework is generic and applicable to the study of properties of quantum phase transitions \cite{Sachdev1997,Hartnoll2007} as well as to quantum spin liquids \cite{spinonhydro2021} or magnon dynamics in ferromagnets \cite{Camilo2019}. 
Hydrodynamic behavior in more conventional electronic solid state systems, such as metals, had been elusive for many decades  \cite{Gurzhi1963}. The main obstacles are usually dominant competing scales due to the underlying lattice of solid-state systems in the form of impurities and lattice vibrations. These limitations have been overcome in recent years, with graphene sitting at the front of this development \cite{Crossno2016,Bandurin2018,Braem2018,Sulpizio2019,Ku2020}.

In the conventional theory of hydrodynamic graphene, there is an underlying assumption: the fluid is composed of electrons and holes that equilibrate locally due to collisions mediated by (long-range) interactions \cite{Kashuba2008,Fritz2008a,Fritz2008b,Fritz2009,Narozhny2015,Lucas2016,Lucas2018,Kiselev2019,kiselev2020,Alekseev2021,Hankiewicz2023}. 
 However, these interactions not only lead to local equilibration, the main assumption underlying hydrodynamics, they also facilitate the emergence of collective excitations, such as charge-density oscillations called plasmons~\cite{PinesBohm1952,BohmPine1953,Vlasov1968}. Those are proper quasiparticles in their own right: In three-dimensional metals, plasmons are inert due to their large excitation gap~\cite{Pines1953}. In two dimensions, however, they follow a square-root dispersion relation~\cite{Stern1967,Wunsch2006,DasSarma2013}. Being gapless, they are important for low-energy equilibrium, for properties such as superconducting instabilities \cite{Takada1978,Lee2017}, and near-equilibrium phenomena \cite{WyldPines1962,Rana2011} such as hydrodynamic transport phenomena  \cite{Schmalian2018,Kitinan2022a,Kitinan2022b,Kitinan2022c}.
	
	In recent works, some of us have studied the thermo-electric response of interacting two-dimensional Dirac systems at and away from the Dirac point. We found a small enhancement of the thermal conductivity due to plasmons at the charge neutrality point. Away from the Dirac point, however, a strong enhancement of the thermal conductivity is found. This can be explained from the undamped gapless nature of plasmons together with their increasing density of states~\cite{Kitinan2022a,Kitinan2022b,Kitinan2022c}. These works suggest that every quantity that depends on energy is sensitive to plasmons. The shear viscosity is directly related to the stress-energy tensor and consequently potentially sees a large plasmon effect. 

 In this work, we study the effect of plasmons on the shear viscosity in interacting graphene or, more generally, two-dimensional Dirac theories at varying fillings (our findings are in fact more generically relevant for two-dimensional Fermi liquids). The shear viscosity of the interacting electron-hole fluid in charge neutral graphene was estimated before in Refs.~\cite{Fritz2009,Link2018}. Now, we go further, including the contribution of the plasmons on equal footing \cite{PinesSchrieffer1962}. A corresponding experiment has recently been carried out not only at the charge-neutrality point but also at moderately high dopings corresponding to the Fermi liquid regime~\cite{Ku2020}. 
 Thus, the goal of this study is twofold: (i) To find an effective hydrodynamic description of a viscous fluid composed of electrons, holes, and plasmons, all on equal footing; (ii) To extend the previous theoretical calculation of the shear viscosity from Ref.~\cite{Fritz2009} away from the Dirac point into the Fermi liquid regime. 
 We use two complementary approaches: an effective Chapman-Enskog approach and a microscopic description in terms of coupled Boltzmann equations for electrons, holes, and plasmons.

Our main result is shown in Fig.~\ref{fig3} which shows the sizeable enhancement of the viscosity due to the plasmons.
 
{\it The model:} We model the non-interacting electronic structure with a low-energy Dirac Hamiltonian of the type $\hat{H}_0 =  \int d\vec{x}\;\hat{\Psi}_{i,\lambda}^\dagger(\vec{x}) \left(-i\hbar v_F \vec{\sigma} \cdot \grad_{\vec{x}}-\mu+V_{\text{ex}}(\vec{x})\right)_{\lambda\lambda'}\hat{\Psi}_{i,\lambda'}(\vec{x})$. The operators $\hat{\Psi}^\dagger_{i,\lambda}(\vec{x})$ and $\hat{\Psi}_{i,\lambda}(\vec{x})$ create and annihilate an electron at a position $\vec{x}$. The effective Fermi velocity is given by $v_F$, approximately $10^6$ m/s for graphene, while the filling is controlled by a chemical potential $\mu$. The static potential $V_{\text{ex}}(\vec{x}) = n_0 \int d\pvec{x}'V(\vec{x}-\vec{x}')$ is added in order to account for the positive charge background of average density $n_0$ in which the electrons move. The flavor index, denoted $i$, ranges from $i=1, ..., N=4$ accounting for spin and valley degrees of freedom (for Dirac systems with a different number of flavors this is an adjustable parameter of the theory). The symbols $\lambda,\lambda'\in \{+,-\}$ denote spinor indices and $\vec{\sigma}=(\sigma_x,\sigma_y)$ are Pauli matrices (note that double indices are summed over) \cite{CastroNeto2009}. 
Additionally, electrons and holes interact via Coulomb interaction according to $\hat{H}_I =1/2  \int d\vec{x}d\vec{x}'\;  \hat{\Psi}_{i,\lambda}^\dagger(\vec{x})\hat{\Psi}_{i,\lambda'}^\dagger(\vec{x}')V(\vec{x}-\vec{x}')\hat{\Psi}_{i,\lambda'}(\vec{x}')\hat{\Psi}_{i,\lambda}(\vec{x})$. The instantaneous Coulomb interaction between two electrons of charge $e$ locating at $\vec{x}$ and $\vec{x}' $ is given by the usual $V(\vec{x}-\vec{x}')=e^2/(4\pi\epsilon|\vec{x}-\vec{x}'|)$, where $\epsilon$ is the average dielectric constant, measuring the average value of the dielectric constant of materials above and below it. The strength of the Coulomb interaction is typically characterized by the ratio of the potential energy to the kinetic energy. This ratio boils down to the fine structure constant $\alpha=e^2/4\pi\epsilon \hbar v_F$ for Dirac systems.  For suspended graphene in vacuum, $\epsilon = 1$  and $\epsilon \approx 7$ for graphene sandwiched between hexagonal boron nitride layers. Thus, for those two cases, $\alpha = 2.2$ and $\alpha \approx 0.3$, respectively \cite{Kotov2012}.

Contrary to Ref.~\cite{Fritz2009}, we go beyond perturbation theory in $\alpha$ and explicitly include collective excitations, most notably plasmons. A key complication in this approach is to find an approximation that respects conservation laws and explicitly does not double count degrees of freedom (cf. Appendix \ref{app:double_count}). 
The basis is the proper use of the random-phase approximation (RPA) which leads to a set of coupled (kinetic) equations of electrons, holes, and plasmons, all on equal footing. The whole procedure is very technical and discussed at length in Refs.~\cite{Kitinan2022a,Kitinan2022b,Kitinan2022c}.

\begin{figure}
    \centering
    \includegraphics[width=\linewidth]{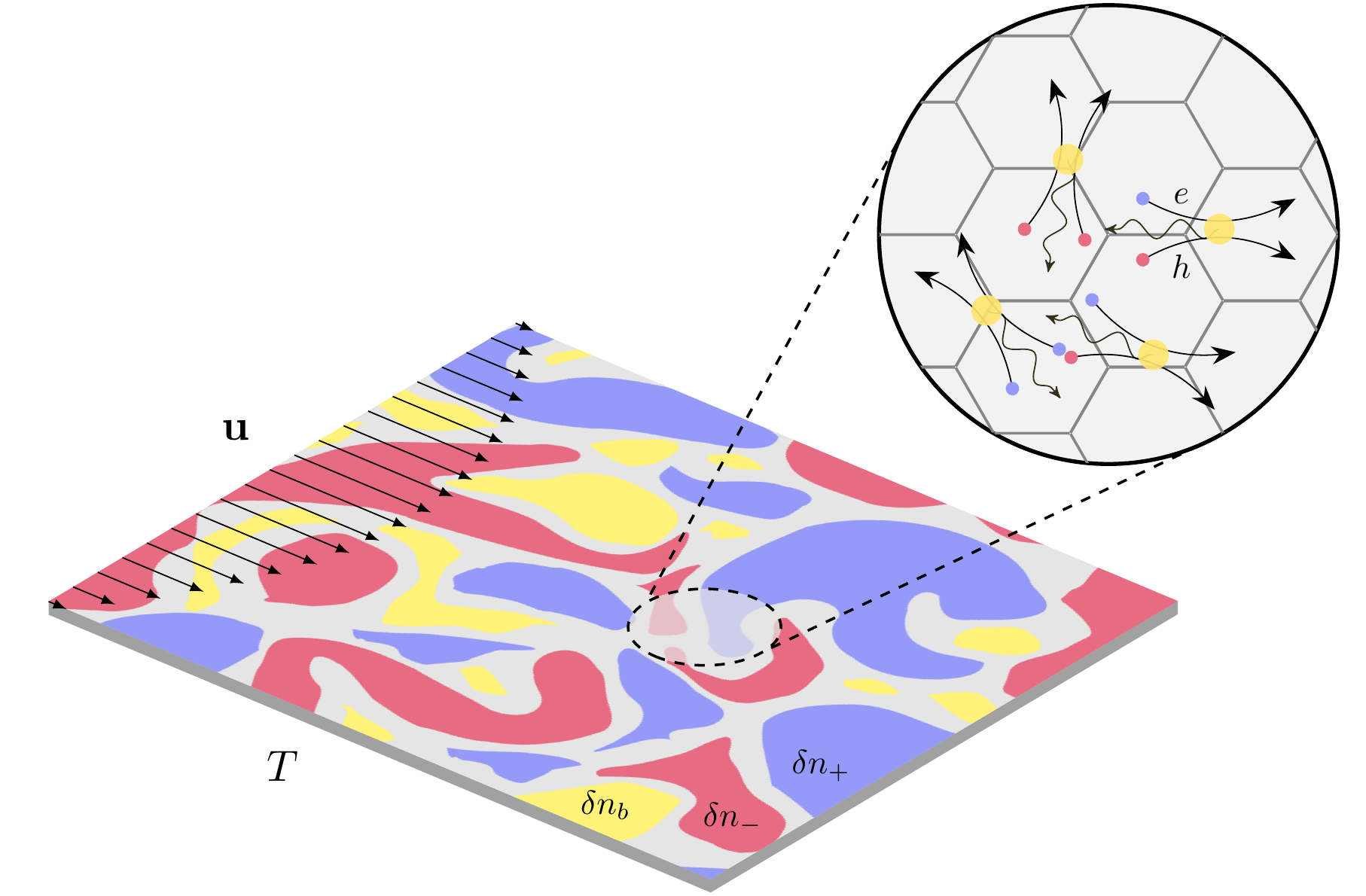}
    \caption{Beyond weak coupling, the fluid of hydrodynamic graphene is composed of three types of particles, electrons, holes, and plasmons; allowing for the local variation of the respective number densities $\delta n_+$, $\delta n_-$, and $\delta n_b$. These particles contribute equally to viscosity, which leads to a parabolic flow profile of the joint velocity $\vec{u}$ through a narrow channel. The three components are coupled via the long-range collisions (cf. Eqs. \eqref{eq:coupledboltzmann}), here schematically represented in the inset.}
    \label{fig1}
\end{figure}

{\it The fluid:} In the hydrodynamic framework of the RPA, the `fluid' is composed of three types of particles, all on equal footing: electrons, holes, and plasmons, see Fig.~\ref{fig1}. 
We introduce an index $\lambda$ that distinguishes electrons ($\lambda=+$) and holes ($\lambda=-$). The respective energies read $\epsilon_{\vec{k},\lambda}=\lambda v_F |\vec{k}|$, and the velocity of a quasiparticle is given by $\vec{v}_{\vec{k},\lambda}=\nabla_{\vec{k}} \epsilon_{\vec{k},\lambda} = \lambda v_F \vec{\hat{k}}$ (note that from this point onwards we set $\hbar=k_B=1$) . We denote the corresponding distribution functions $f_\lambda(\vec{k},\vec{x},t)$ that reduce to the Fermi function in thermal equilibrium.

For the plasmons, we review the most salient features here, more details can be found in Ref.~\cite{Kitinan2022a,Kitinan2022b,Kitinan2022c} and cited papers. The plasmon dispersion relation is given by $\omega_{\vec{q}} = \sqrt{\frac{N}{2}\alpha v_F T  \log(2+2\cosh(\mu/T))q}\equiv\sqrt{\mathcal{N}q}$ where $T$ is the temperature and $q$ is the modulus of two-dimensional momentum. The velocity of the plasmons is given by $\vec{w}_{\vec{q}}=\nabla_\vec{q}\omega(\vec{q})$. The requirement for a Boltzmann equation to be valid is the existence of well-defined quasi-particle excitation. A plasmon of high momentum is overly damped into a particle-hole continuum, so only the plasmon of small momentum, or long wavelength, are well-defined. Consequently, there is a momentum cutoff for plasmons beyond which their spectral function is broadened significantly. The momentum cutoff is found to be $q_c=\mathcal{N}$~\cite{Kitinan2022a}. The distribution function of the plasmons is henceforth called $b(\vec{q},\vec{x},t)$.

Although the fluid is composed of three distinct species with differing quasiparticle velocities, all the components are locked in a uniform flow of velocity $\vec{u}$. This locking is a consequence of entropic constraints, as shown in Appendix~\ref{app:htheorem}. Therefore, the viscous forces from each sector -- be it fermions or plasmons -- arise from the same velocity field. 

{\it Effective Boltzmann equations:} Formally, the Boltzmann equations (BE) can be derived from the Keldysh formalism. Here, however, we content ourselves with using them and refer the reader to Ref.~\cite{Kitinan2022a,Kitinan2022b,Kitinan2022c}. We find three coupled equations:
\begin{eqnarray}\label{eq:coupledboltzmann}
\partial_t f_\lambda+ \vec{v}_{\vec{k},\lambda}\cdot \nabla_{\vec{x}} f_\lambda -\nabla_{\vec{x}}\epsilon_{\vec{k},\lambda}\cdot\nabla_{\vec{k}} f_\lambda &=&\mathcal{I}^f_{\lambda}[f,b] \;,\nonumber \\ \partial_t b+\vec{w}_{\vec{q}}\cdot \grad_{\vec{x}} b -\nabla_{\vec{x}}\omega_{\vec{q}}\cdot\nabla_{\vec{q}} b&=& \mathcal{I}^b[f,b]\;.
\end{eqnarray}
The left hand side contains the so-called streaming terms accounting for spatial and temporal variations. The right hand side contains the collision terms, $\mathcal{I}^b[f,b]$ and $\mathcal{I}^f_{\lambda}[f,b]$, that couple the three equations explicitly. A detailed discussion of the scattering terms has been relegated to Appendix~\ref{app:collisions}. It is important to note that for the problem at hand we do not require a translational symmetry-breaking collision term for the reason that viscosity manifests itself in a situation in which translational symmetry is broken by boundary conditions. 


Performing momentum integrals with respect to the moments of the distribution function allows to derive the conservation laws. In the present work we are considering the first two moments, i.e. $\int_{\vec{k}} \delta f_{\lambda,\vec{k}}$ and $\int_{\vec{k}} \vec{k}\delta f_{\lambda,\vec{k}}$, that state conservation of particle number and momentum; in particular, we can evaluate the momentum flux, which comprises the average momentum current $\vec{u}\vec{\bar{p}}$, pressure $P$, and viscous terms as
$\mtx{\Pi}=\vec{u}\vec{\bar{p}}+\mtx{1}P+\delta\mtx{\Pi}$, where the last term corresponds to the viscous contribution. Decomposing the distribution functions into their equilibrium and fluctuation correction contributions, {\it i.e.},  $f_{\lambda} = f^0_{\lambda,\vec{k}}+\delta f_{\lambda,\vec{k}}$ and $b = b^0_\vec{q}+\delta b_\vec{q}$, allow us to write  
\begin{equation}
\label{eq:momentumflux}
 \delta \mtx{\Pi} = \sum_{\lambda} N\!\! \int\!\! \frac{d\vec{k}}{(2\pi)^2} \vec{v}_{\lambda,\vec{k}}\; \vec{k}\; \delta f_{\lambda,\vec{k}} + \!\!\int\!\! \frac{d\vec{q}}{(2\pi)^2} \vec{w}_{\vec{q}}\vec{q}\delta b_\vec{q}.
\end{equation}

We use two complementary methods to account for the fluctuations of the distribution functions coming from the collision operators: (a) the Chapman--Enskog method and (b) the linear response method, including a numerical solution of the BEs.

{\it The Chapman-Enskog approach:} As a first approach to calculate the viscosity of the three component fluid, we resort to the Chapman--Enskog (CE) method~\cite{Chapman1954TheGases} with a Bhatnagar--Gross--Krook (BGK) approximation~\cite{PhysRev.94.511,Gross1956} for the collision operators. This is a way to effectively decouple the equations using the so-called relaxation time approximation:
\begin{equation}
    \delta f_{\lambda,\vec{k}}=\tau_\lambda\mathfrak{D}f^0_\lambda\quad\text{and} \quad    \delta b_{\vec{q}}=\tau_b\mathfrak{D}b^0 \label{eq:CEleading}\;.
\end{equation} 
This is the leading order CE expansion, where the differential operator $\mathfrak{D}\doteq\partial_t+ \vec{v}_{\vec{k},\lambda}\cdot \nabla_{\vec{x}} -\nabla_{\vec{x}}\epsilon_{\vec{k},\lambda}\cdot\nabla_{\vec{k}}$ is shorthand for the LHS of Boltzmann equations, Eq.~\eqref{eq:coupledboltzmann}, and the relaxation times
$\tau_\lambda$ and $\tau_b$ combine the collision mechanisms providing an effective phenomenological description of the problem \cite{schuett2011,Fu2018}.  



The velocity of electrons (holes) with respect to the fluid velocity reads $\vec{c}_\lambda=\lambda v_F\vec{\hat{k}}-\vec{u}$, which allows to rewrite the streaming term of the Boltzmann operator as 
\begin{multline}
\mathfrak{D}f_\lambda^0=
    f_\lambda^0(1-f_\lambda^0)\Bigg\{%
    \frac{\vec{c}_\lambda\cdot \grad T}{T}\left(\frac{\vec{c}_\lambda\cdot\vec{k}}{T}-\frac{3P_\lambda}{n T}\right)+\\+
    \frac{\vec{k}}{T}\cdot\left[(\vec{c}_\lambda \cdot\grad)\vec{u}-\frac{\vec{c}_\lambda{-\vec{u}}}{2}\grad\cdot \vec{u}\right]
    \Bigg\}
\end{multline}
where we made use of the Euler fluid equations to eliminate the time derivatives. 
%
%
%
%
%
This leads to a viscous tensor of the form 
\begin{equation}
    \delta\mtx{\Pi}_\lambda= -\frac{3\tau_\lambda P_\lambda}{4}
  \left(2\dot{\mtx{e}}-\mtx{1}\Tr \dot{\mtx{e}}\right)
\end{equation}
where $\dot e_{ij}=(\partial_iu_j+\partial_ju_i)/2$ is the strain-rate tensor. The relaxation time $\tau$ shown here is to be suitably chosen following Eq.\eqref{eq:CEleading} and it is assumed to have no dependence on the momentum as first approximation. Thus, the shear viscosity due to the fermions can be identified as $\eta=\sum_\lambda3\tau_\lambda P_\lambda/4$. Note that there is no contribution of the bulk viscosity, as expected for systems with linear dispersion \cite{Gabbana2020RelativisticApplications}. Moreover, the Fermi liquid limit of this result agrees with the recent literature \cite{Chen2022ViscosityElectrons}. Additionally, if one wishes to account for the fermionic self-energy contributions this can easily be achieved by accounting for renormalisation effects via the Fermi velocity. However, such corrections have a negligible impact on the global behavior of the viscosity (see, e.g., Appendix \ref{app:vfrenormalization}).


For the plasmons, we start with the aforementioned square root dispersion relation $\omega(q)=\sqrt{\mathcal{N}q}$, 
The relative velocity is defined as $\vec{C}=\partial_q\omega\vec{\hat{q}}-\vec{u}$, implying that $\omega(q)=2(\vec{C}+\vec{u})\cdot \vec{q}$. This treatment makes the {\it a priori} unjustified assumption that the fluid of plasmons flows at the same velocity as the fluid of the electrons and holes. We show in Appendix~\ref{app:htheorem} that this is not an assumption but can in fact be derived from the underlying BE. The streaming term now reads
\begin{multline}
    \mathfrak{D}b^0=b^0(1+b^0)\Bigg\{  \frac{\vec{q}\cdot(2\vec{C}+\vec{u})}{T^2}\vec{C}\cdot\grad T +\\+\frac{\vec{q}}{T}\cdot\left[(\vec{C}\cdot\grad)\vec{u}-\frac{3}{5}\vec{C}\diver\vec{u}+\frac{9}{20}\vec{u}\diver \vec{u}\right]  \Bigg\}.
\end{multline}

\begin{figure}[ht!]
    \centering
    \includegraphics[width=.9\linewidth]{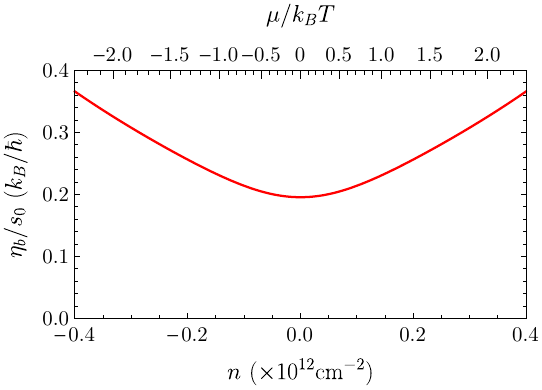}
    \caption{Plasmonic contribution to viscosity $\alpha=0.9$ evaluated at $T=300\,\mathrm{K}$ from Eq. \eqref{eq:visco_plasmons}.}
    \label{fig:visco_plasmons}
\end{figure}

\begin{figure}[tbh]
		\centering
		\includegraphics[width=.9\linewidth]{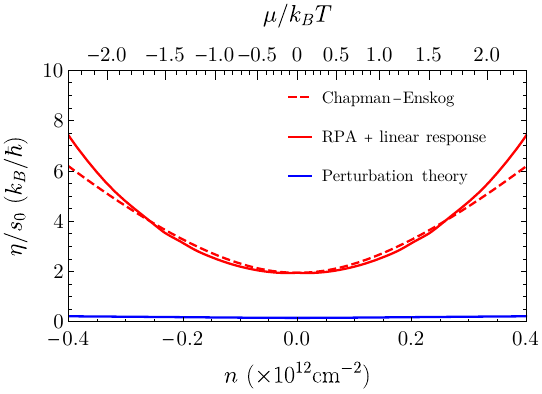}
		\caption{The ratio of shear viscosity $\eta$ to $s_0$ evaluated at the room temperature $T =300$ K as a function of charge density $n$. Here $\eta/s_0$ is plotted in the unit of $k_B/\hbar$ where $s_0$ is electron entropy density at the Dirac point. In this plot, the fine-structure constant $\alpha \approx 0.9$. } 
		\label{fig3}
\end{figure}

Choosing the correct form of the relaxation time for the plasmons is more delicate and in principle relies on a microscopic treatment of the coupled equations. However, since the coupled equations can be understood as a wave-particle interaction, it is natural to consider Landau damping as the main relaxation mechanism. From RPA calculations \cite{Kitinan2022b,Kitinan2022c} the decay rate is given by  
\begin{equation}
    \frac{1}{\tau_b} = -\frac{\pi \omega(q)^2\big[n_f(\omega(q)/2)-n_f(-\omega(q)/2)\big]}{8T \log\left(2+2\cosh \mu/T\right)}\label{eq:landaudamping}
\end{equation}
where $n_f(\cdot)$ is the equilibrium Fermi-Dirac distribution. We used Eq,~\eqref{eq:landaudamping} to compute the bosonic contribution to viscosity. The viscous tensor is given by 
\begin{equation}
    \delta\mtx{\Pi}_b = -\eta_b\left[2\dot{\mtx{e}}-\mtx{1}\Tr \dot{\mtx{e}}   -\mtx{1}\frac{\diver\vec{u}}{10}\right]
\end{equation}
where the viscosity is given by the integral 
\begin{equation}
    \eta_b= \bigintssss_0^{q_c}\!\! \frac{d}{dq}\left(\frac{\mathcal{N}^2\tau_b(q)}{2}q^{5/2}\right)b^0\,dq\;.\label{eq:visco_plasmons}
\end{equation}
We limit the integration to the cut-off at $q_c$ in order to ensure that the contribution of plasmons is bounded within the range where they are long-lived. The evaluation of the shear viscosity, arising from the plasmon sector, is showcased in   Fig.\ref{fig:visco_plasmons}.
Note that, contrary to the case of fermions, when taking the divergence of this tensor one will get both shear viscosity and bulk viscosity $\diver\mtx{\Pi}_b=-\eta_b\Big[\nabla^2\vec{u}-\frac{1}{10}\grad(\diver\vec{u})\Big]$, the latter being, however, small compared to the shear part. 

We conclude that there is an effect of the plasmons the total viscous forces. 

While the relaxation time approximation is very useful to understand qualitative features of the system, it does not allow for quantitative statements, because it relies on a particular choice of the relaxation time value and behavior. Therefore, we will now devise a strategy to determine the relaxation time based on a linear-response theory solution of the Boltzmann equation.

{\it Boltzmann equation and linear response theory:} The BGK procedure has the relaxation time as a free parameter. A technique complementary to the CE procedure is to solve the coupled Boltzmann equations, Eqs.~\eqref{eq:coupledboltzmann}, within linear response. This gives direct access to the relaxation times from a microscopic theory. The collision integrals $\mathcal{I}^b[f,b]$ and $\mathcal{I}^f_{\lambda}[f,b]$ describe collision processes involving two fermions and a single plasmon and are detailed in Appendix~\ref{app:collisions} with their corresponding Feynman diagrams.
%
In the presence of a small velocity gradient, the streaming terms on the left-hand-side assume the form
	$\vec{w}_{\vec{q}} \!\cdot\! \grad_{\vec{x}} b^0_\vec{q} = \frac{\omega_{\vec{q}}}{2T} b^0_\vec{q}\left(1+b^0_\vec{q}\right) \left[\left(\hat{q}_i\hat{q}_j-\delta_{ij}\right)\frac{X_{ij}}{2} +\frac{\diver\vec{u}}{2}\right]$ and $\vec{v}_{\lambda,\vec{k}} \!\cdot\!\grad_{\vec{x}} f^0_{\lambda,\vec{k}} = \frac{\epsilon_{\lambda,\vec{k}}}{T} f_{\lambda,\vec{k}}^0\left(1-f_{\lambda,\vec{k}}^0\right) \left[\left(\hat{k}_i\hat{k}_j-\delta_{ij}\right)\frac{X_{ij}}{2} +\frac{\diver\vec{u}}{2}\right]$
in linear response where we introduced $\mtx{X}_{ij} =2\dot{\mtx{e}}-\mtx{1}\Tr \dot{\mtx{e}}  $.
Let us note that the last term in the square bracket proportional to $\diver\vec{u}$ is responsible for the bulk viscosity, which we disregard from now on\footnote{As we have seen the CE method shows that the bulk viscosity of the fermionic sector should be identically null and that the for the plasmon sector it is $\sim1/10$ of the shear counterpart.}.
The driving terms motivate a parametrization of the deviation of the distribution functions from their equilibrium values $f^0_{\lambda}(\vec{k})$ and $b^0(\vec{q})$ in the standard way. We write $f_{\lambda}(\vec{x},\vec{k},t) = f^0_\lambda(\vec{k})+\delta f_{\lambda}(\vec{k})$ and $b(\vec{q},\vec{x},t) = b^0(\vec{q})+\delta b(\vec{q})$, where

\begin{equation}
	\label{eq:parbos}
	\begin{split}
	\delta b(\vec{q}) &= \frac{g^b(\vec{q})}{T}\frac{\omega_{\vec{q}}}{2T} b^0(\vec{q})\left(1+b^0(\vec{q})\right) \sum_{ij}\left(\frac{\vec{q}_i\vec{q}_j}{q^2}-\vec{\vec{1}}_{ij}\right)\vec{\vec{X}}_{ij}, \\ &\equiv b^0(\vec{q})\left(1+b^0(\vec{q})\right)\sum_{ij}g^b_{ij}(\vec{q})\vec{\vec{X}}_{ij},
	\end{split}
\end{equation}
and 
\begin{equation}
\label{eq:parferm}
	\begin{split}
	&\delta f_{\lambda}(\vec{k}) \\&\;\;= \frac{g^f_{\lambda}(\vec{k})}{T}\frac{\epsilon_{\lambda,\vec{k}}}{T} f_\lambda^0(\vec{k})\left(1-f_\lambda^0(\vec{k})\right) \sum_{ij}\left(\frac{\vec{k}_i\vec{k}_j}{k^2}-\vec{\vec{1}}_{ij}\right)\vec{\vec{X}}_{ij}, \\ &\;\;\equiv f_\lambda^0(\vec{k})\left(1-f_\lambda^0(\vec{k})\right)\sum_{ij}g^f_{\lambda,ij}(\vec{k})\vec{\vec{X}}_{ij}. 
	\end{split}
\end{equation}
We proceed to the linearization of the collision integrals by inserting the parametrization \eqref{eq:parbos} and \eqref{eq:parferm} into \eqref{eq:colbos} and \eqref{eq:colferm} and find that
\begin{widetext}
	\begin{equation}
		\mathcal{I}_{(1)}^b[f,b]=-N\sum_{\lambda,\lambda'}\int \frac{d\vec{k}}{(2\pi)^2}  \mathcal{M}_{\lambda\lambda'}^{\vec{k}+\vec{q},\vec{k}}\delta(\omega_{\vec{q}}+\epsilon_{\lambda,\vec{k}}-\epsilon_{\lambda',\vec{k}+\vec{q}} )\left[f^0_\lambda(\vec{k})\left(1-f^0_{\lambda'}(\vec{k}+\vec{q})\right)b^0(\vec{q})\right]\left[ g^{f}_{\lambda,ij}(\vec{k})-g^f_{\lambda',ij}(\vec{k}+\vec{q})+g^b_{ij}(\vec{q})\right]\vec{\vec{X}}_{ij},
	\end{equation}
	\begin{equation}
	\begin{split}
		\mathcal{I}^f_{(1)\lambda}[f,b]&=-\sum_{\lambda'}\int \frac{d\vec{q}}{(2\pi)^2}  \mathcal{M}_{\lambda\lambda'}^{\vec{k}+\vec{q},\vec{k}}\delta(\omega_{\vec{q}}+\epsilon_{\lambda,\vec{k}}-\epsilon_{\lambda',\vec{k}+\vec{q}} )\left[f^0_\lambda(\vec{k})\left(1-f^0_{\lambda'}(\vec{k}+\vec{q})\right)b^0(\vec{q})  \right]\left[ g^{f}_{\lambda,ij}(\vec{k})-g^f_{\lambda',ij}(\vec{k}+\vec{q})+g^b_{ij}(\vec{q})\right]\vec{\vec{X}}_{ij}\\
		&~-\sum_{\lambda'}\int \frac{d\vec{q}}{(2\pi)^2}  \mathcal{M}_{\lambda\lambda'}^{\vec{k}-\vec{q},\vec{k}}\delta(-\omega_{\vec{q}}+\epsilon_{\lambda,\vec{k}}-\epsilon_{\lambda',\vec{k}-\vec{q}} )\left[\left(1-f^0_\lambda(\vec{k})\right)f^0_{\lambda'}(\vec{k}-\vec{q})b^0(\vec{q}) \right] \left[ g^{f}_{\lambda',ij}(\vec{k}-\vec{q})-g^f_{\lambda,ij}(\vec{k})+g^b_{ij}(\vec{q})\right]\vec{\vec{X}}_{ij}.
	\end{split}
	\end{equation}
\end{widetext}
The remainder is to find the solution of the linearized Boltzmann equations for the function $g_{ij}^b(\vec{q})$ and $g^{f}_{\lambda,ij}(\vec{k})$ by means of a variational method. This method is standard and has been intensively used for solving a Boltzmann equation (see, for example, Ref.~\cite{Ziman2000}).

{\it{Results:}} For a slowly varying velocity field, the correction to the stress tensor can be expanded in gradients of $\vec{u}(\vec{x})$. The shear viscosity characterizes the diffusive relaxation of transverse momentum density fluctuations measuring the resistance of a fluid to establish a transverse velocity gradient. 
In the model of the three-component fluid of electrons, holes, and plasmons, Eq.~\eqref{eq:momentumflux} describes the momentum flux which can be related to the shear viscosity according to $\delta \Pi_{ij} = -\eta_{ijkl} \dot{e}_{kl}$. Microscopically, this leads to
\begin{widetext}
\begin{equation}
\eta_{ijkl} = \frac{1}{8}\left(\sum_{\lambda}N\!\!\int\!\!\frac{d\vec{k}}{(2\pi)^2} \frac{(\epsilon_{\lambda,\vec{k}})^2}{T^2}\;g^f_\lambda(\vec{k}) f_{\lambda,\vec{k}}^0(1-f_{\lambda,\vec{k}}^0)+\!\!\int\!\!\frac{d\vec{q}}{(2\pi)^2}\frac{(\omega_{\vec{q}})^2}{4T^2}g_b(\vec{q})b^0_\vec{q}(1+b^0_\vec{q})\right)\left(\delta_{ik}\delta_{jl}+\delta_{il}\delta_{jk}-\delta_{ij}\delta_{kl}\right).
\end{equation}
\end{widetext}
which is what we calculated numerically. In order to estimate the size of the effect of collective plasmon modes, we also calculated the viscosity in perturbation theory following Ref.~\cite{Fritz2009}. For a value of $\alpha\approx 0.9$, the result is plotted in Fig.~\ref{fig3}. As a function of the density, the red curve shows the total viscosity including fermions and plasmons, whereas the blue curve shows the viscosity of the fermions only calculated in perturbation theory in $\alpha$. We find that there is a drastic increase in viscous effects due to plasmons.

{\it{Conclusions:}} In this work, we computed the contribution to the shear viscosity from the plasmons in the charge fluid of graphene. We employed a multi-fluid model describing electrons, holes and plasmons in line with Refs. \cite{Kitinan2022a,Kitinan2022b,Kitinan2022c}. Following both relaxation time approximation and linear response theory methods, we arrive to agreeing results that show a possibly strong enhancement of the viscosity due to the electron/hole-plasmon interactions. Comparing our results to experiment will shed new light on the interaction physics of graphene and might allow for an alternative determination of the fine structure constant. Beyond graphene, we believe that our result is generic to two-dimensional metals which will allow for new insights into the physics of two-dimensional Fermi liquids \cite{Galitski2020,zverevich2023transport}.

\textit{Acknowledgements:}
K.P. and L.F. acknowledge former collaborations and discussions with H.T.C. Stoof, T. Ludwig, S. Grubinskas, R.A. Duine, E.I. Kiselev, A. Lucas, J. Lux, A. Mitchell, M. M\"uller, A. Rosch, S. Sachdev, J. Schmalian, H. Terças, M. Vojta, G. Wagner, T. Meng, and J. Waissman. KP thanks the Institute for the Promotion of Teaching Science and Technology (IPST) of Thailand for a Ph.D. scholarship. One of the authors (P.C.) acknowledges the funding provided by Fundação para a Ciência e a Tecnologia (FCT-Portugal) through the Grant No. PD/BD/150415/2019 and Project No. PTDC/FIS-OUT/3882/2020. This work is part of the D-ITP consortium, a program of the Netherlands Organisation for Scientific Research (NWO) that is funded by the Dutch Ministry of Education, Culture and Science (OCW).

K.P. and P.C. contributed equally to this work.
\bibliographystyle{apsrev4-2}


\bibliography{referencesviscosity}

\begin{thebibliography}{62}%
\makeatletter
\providecommand \@ifxundefined [1]{%
 \@ifx{#1\undefined}
}%
\providecommand \@ifnum [1]{%
 \ifnum #1\expandafter \@firstoftwo
 \else \expandafter \@secondoftwo
 \fi
}%
\providecommand \@ifx [1]{%
 \ifx #1\expandafter \@firstoftwo
 \else \expandafter \@secondoftwo
 \fi
}%
\providecommand \natexlab [1]{#1}%
\providecommand \enquote  [1]{``#1''}%
\providecommand \bibnamefont  [1]{#1}%
\providecommand \bibfnamefont [1]{#1}%
\providecommand \citenamefont [1]{#1}%
\providecommand \href@noop [0]{\@secondoftwo}%
\providecommand \href [0]{\begingroup \@sanitize@url \@href}%
\providecommand \@href[1]{\@@startlink{#1}\@@href}%
\providecommand \@@href[1]{\endgroup#1\@@endlink}%
\providecommand \@sanitize@url [0]{\catcode `\\12\catcode `\$12\catcode
  `\&12\catcode `\#12\catcode `\^12\catcode `\_12\catcode `\%12\relax}%
\providecommand \@@startlink[1]{}%
\providecommand \@@endlink[0]{}%
\providecommand \url  [0]{\begingroup\@sanitize@url \@url }%
\providecommand \@url [1]{\endgroup\@href {#1}{\urlprefix }}%
\providecommand \urlprefix  [0]{URL }%
\providecommand \Eprint [0]{\href }%
\providecommand \doibase [0]{https://doi.org/}%
\providecommand \selectlanguage [0]{\@gobble}%
\providecommand \bibinfo  [0]{\@secondoftwo}%
\providecommand \bibfield  [0]{\@secondoftwo}%
\providecommand \translation [1]{[#1]}%
\providecommand \BibitemOpen [0]{}%
\providecommand \bibitemStop [0]{}%
\providecommand \bibitemNoStop [0]{.\EOS\space}%
\providecommand \EOS [0]{\spacefactor3000\relax}%
\providecommand \BibitemShut  [1]{\csname bibitem#1\endcsname}%
\let\auto@bib@innerbib\@empty
\bibitem [{\citenamefont {Landau}\ and\ \citenamefont
  {Lifshitz}(1987)}]{LandauLifshitz1987}%
  \BibitemOpen
  \bibfield  {author} {\bibinfo {author} {\bibfnamefont {L.}~\bibnamefont
  {Landau}}\ and\ \bibinfo {author} {\bibfnamefont {E.}~\bibnamefont
  {Lifshitz}},\ }\href@noop {} {\emph {\bibinfo {title} {Fluid Mechanics:
  Course of Theoretical Physics}}}\ (\bibinfo  {publisher} {Pergamon Press},\
  \bibinfo {year} {1987})\BibitemShut {NoStop}%
\bibitem [{\citenamefont {Hartnoll}\ \emph {et~al.}(2007)\citenamefont
  {Hartnoll}, \citenamefont {Kovtun}, \citenamefont {M\"uller},\ and\
  \citenamefont {Sachdev}}]{Hartnoll2007}%
  \BibitemOpen
  \bibfield  {author} {\bibinfo {author} {\bibfnamefont {S.~A.}\ \bibnamefont
  {Hartnoll}}, \bibinfo {author} {\bibfnamefont {P.~K.}\ \bibnamefont
  {Kovtun}}, \bibinfo {author} {\bibfnamefont {M.}~\bibnamefont {M\"uller}},\
  and\ \bibinfo {author} {\bibfnamefont {S.}~\bibnamefont {Sachdev}},\ }\href
  {https://doi.org/10.1103/PhysRevB.76.144502} {\bibfield  {journal} {\bibinfo
  {journal} {Phys. Rev. B}\ }\textbf {\bibinfo {volume} {76}},\ \bibinfo
  {pages} {144502} (\bibinfo {year} {2007})}\BibitemShut {NoStop}%
\bibitem [{\citenamefont {Andreev}\ \emph {et~al.}(2011)\citenamefont
  {Andreev}, \citenamefont {Kivelson},\ and\ \citenamefont
  {Spivak}}]{Spivak2011}%
  \BibitemOpen
  \bibfield  {author} {\bibinfo {author} {\bibfnamefont {A.~V.}\ \bibnamefont
  {Andreev}}, \bibinfo {author} {\bibfnamefont {S.~A.}\ \bibnamefont
  {Kivelson}},\ and\ \bibinfo {author} {\bibfnamefont {B.}~\bibnamefont
  {Spivak}},\ }\href {https://doi.org/10.1103/PhysRevLett.106.256804}
  {\bibfield  {journal} {\bibinfo  {journal} {Phys. Rev. Lett.}\ }\textbf
  {\bibinfo {volume} {106}},\ \bibinfo {pages} {256804} (\bibinfo {year}
  {2011})}\BibitemShut {NoStop}%
\bibitem [{\citenamefont {Zaanen}(2016)}]{Zaanen2016}%
  \BibitemOpen
  \bibfield  {author} {\bibinfo {author} {\bibfnamefont {J.}~\bibnamefont
  {Zaanen}},\ }\href {https://doi.org/10.1126/science.aaf2487} {\bibfield
  {journal} {\bibinfo  {journal} {Science}\ }\textbf {\bibinfo {volume}
  {351}},\ \bibinfo {pages} {1026} (\bibinfo {year} {2016})},\ \Eprint
  {https://arxiv.org/abs/https://www.science.org/doi/pdf/10.1126/science.aaf2487}
  {https://www.science.org/doi/pdf/10.1126/science.aaf2487} \BibitemShut
  {NoStop}%
\bibitem [{\citenamefont {Levitov}\ and\ \citenamefont
  {Falkovich}(2016)}]{Levitov2016}%
  \BibitemOpen
  \bibfield  {author} {\bibinfo {author} {\bibfnamefont {L.}~\bibnamefont
  {Levitov}}\ and\ \bibinfo {author} {\bibfnamefont {G.}~\bibnamefont
  {Falkovich}},\ }\href {https://doi.org/10.1038/nphys3667} {\bibfield
  {journal} {\bibinfo  {journal} {Nature Physics}\ }\textbf {\bibinfo {volume}
  {12}},\ \bibinfo {pages} {672} (\bibinfo {year} {2016})}\BibitemShut
  {NoStop}%
\bibitem [{\citenamefont {Lucas}\ and\ \citenamefont {Fong}(2018)}]{Lucas2018}%
  \BibitemOpen
  \bibfield  {author} {\bibinfo {author} {\bibfnamefont {A.}~\bibnamefont
  {Lucas}}\ and\ \bibinfo {author} {\bibfnamefont {K.~C.}\ \bibnamefont
  {Fong}},\ }\href {https://doi.org/10.1088/1361-648x/aaa274} {\bibfield
  {journal} {\bibinfo  {journal} {Journal of Physics: Condensed Matter}\
  }\textbf {\bibinfo {volume} {30}},\ \bibinfo {pages} {053001} (\bibinfo
  {year} {2018})}\BibitemShut {NoStop}%
\bibitem [{\citenamefont {Narozhny}(2019)}]{Narozhny2019}%
  \BibitemOpen
  \bibfield  {author} {\bibinfo {author} {\bibfnamefont {B.~N.}\ \bibnamefont
  {Narozhny}},\ }\href
  {https://doi.org/https://doi.org/10.1016/j.aop.2019.167979} {\bibfield
  {journal} {\bibinfo  {journal} {Annals of Physics}\ }\textbf {\bibinfo
  {volume} {411}},\ \bibinfo {pages} {167979} (\bibinfo {year}
  {2019})}\BibitemShut {NoStop}%
\bibitem [{\citenamefont {Fritz}\ and\ \citenamefont
  {Scaffidi}(2023)}]{fritz2023hydrodynamic}%
  \BibitemOpen
  \bibfield  {author} {\bibinfo {author} {\bibfnamefont {L.}~\bibnamefont
  {Fritz}}\ and\ \bibinfo {author} {\bibfnamefont {T.}~\bibnamefont
  {Scaffidi}},\ }\href@noop {} {\bibinfo {title} {Hydrodynamic electronic
  transport}} (\bibinfo {year} {2023}),\ \Eprint
  {https://arxiv.org/abs/2303.14205} {arXiv:2303.14205 [cond-mat.str-el]}
  \BibitemShut {NoStop}%
\bibitem [{\citenamefont {Kovtun}(2012)}]{Kovtun2012}%
  \BibitemOpen
  \bibfield  {author} {\bibinfo {author} {\bibfnamefont {P.}~\bibnamefont
  {Kovtun}},\ }\href {https://doi.org/10.1088/1751-8113/45/47/473001}
  {\bibfield  {journal} {\bibinfo  {journal} {Journal of Physics A:
  Mathematical and Theoretical}\ }\textbf {\bibinfo {volume} {45}},\ \bibinfo
  {pages} {473001} (\bibinfo {year} {2012})}\BibitemShut {NoStop}%
\bibitem [{\citenamefont {Damle}\ and\ \citenamefont
  {Sachdev}(1997)}]{Sachdev1997}%
  \BibitemOpen
  \bibfield  {author} {\bibinfo {author} {\bibfnamefont {K.}~\bibnamefont
  {Damle}}\ and\ \bibinfo {author} {\bibfnamefont {S.}~\bibnamefont
  {Sachdev}},\ }\href {https://doi.org/10.1103/PhysRevB.56.8714} {\bibfield
  {journal} {\bibinfo  {journal} {Phys. Rev. B}\ }\textbf {\bibinfo {volume}
  {56}},\ \bibinfo {pages} {8714} (\bibinfo {year} {1997})}\BibitemShut
  {NoStop}%
\bibitem [{\citenamefont {Bulchandani}\ \emph {et~al.}(2021)\citenamefont
  {Bulchandani}, \citenamefont {Hsu}, \citenamefont {Herzog},\ and\
  \citenamefont {Sondhi}}]{spinonhydro2021}%
  \BibitemOpen
  \bibfield  {author} {\bibinfo {author} {\bibfnamefont {V.~B.}\ \bibnamefont
  {Bulchandani}}, \bibinfo {author} {\bibfnamefont {B.}~\bibnamefont {Hsu}},
  \bibinfo {author} {\bibfnamefont {C.~P.}\ \bibnamefont {Herzog}},\ and\
  \bibinfo {author} {\bibfnamefont {S.~L.}\ \bibnamefont {Sondhi}},\ }\href
  {https://doi.org/10.1103/PhysRevB.104.235412} {\bibfield  {journal} {\bibinfo
   {journal} {Phys. Rev. B}\ }\textbf {\bibinfo {volume} {104}},\ \bibinfo
  {pages} {235412} (\bibinfo {year} {2021})}\BibitemShut {NoStop}%
\bibitem [{\citenamefont {Ulloa}\ \emph {et~al.}(2019)\citenamefont {Ulloa},
  \citenamefont {Tomadin}, \citenamefont {Shan}, \citenamefont {Polini},
  \citenamefont {van Wees},\ and\ \citenamefont {Duine}}]{Camilo2019}%
  \BibitemOpen
  \bibfield  {author} {\bibinfo {author} {\bibfnamefont {C.}~\bibnamefont
  {Ulloa}}, \bibinfo {author} {\bibfnamefont {A.}~\bibnamefont {Tomadin}},
  \bibinfo {author} {\bibfnamefont {J.}~\bibnamefont {Shan}}, \bibinfo {author}
  {\bibfnamefont {M.}~\bibnamefont {Polini}}, \bibinfo {author} {\bibfnamefont
  {B.~J.}\ \bibnamefont {van Wees}},\ and\ \bibinfo {author} {\bibfnamefont
  {R.~A.}\ \bibnamefont {Duine}},\ }\href
  {https://doi.org/10.1103/PhysRevLett.123.117203} {\bibfield  {journal}
  {\bibinfo  {journal} {Phys. Rev. Lett.}\ }\textbf {\bibinfo {volume} {123}},\
  \bibinfo {pages} {117203} (\bibinfo {year} {2019})}\BibitemShut {NoStop}%
\bibitem [{\citenamefont {Gurzhi}(1963)}]{Gurzhi1963}%
  \BibitemOpen
  \bibfield  {author} {\bibinfo {author} {\bibfnamefont {R.}~\bibnamefont
  {Gurzhi}},\ }\href@noop {} {\bibfield  {journal} {\bibinfo  {journal}
  {J.Exptl. Theoret.Phys.(U.S.S.R.)}\ ,\ \bibinfo {pages} {771}} (\bibinfo
  {year} {1963})}\BibitemShut {NoStop}%
\bibitem [{\citenamefont {Crossno}\ \emph {et~al.}(2016)\citenamefont
  {Crossno}, \citenamefont {Shi}, \citenamefont {Wang}, \citenamefont {Liu},
  \citenamefont {Harzheim}, \citenamefont {Lucas}, \citenamefont {Sachdev},
  \citenamefont {Kim}, \citenamefont {Taniguchi}, \citenamefont {Watanabe},
  \citenamefont {Ohki},\ and\ \citenamefont {Fong}}]{Crossno2016}%
  \BibitemOpen
  \bibfield  {author} {\bibinfo {author} {\bibfnamefont {J.}~\bibnamefont
  {Crossno}}, \bibinfo {author} {\bibfnamefont {J.~K.}\ \bibnamefont {Shi}},
  \bibinfo {author} {\bibfnamefont {K.}~\bibnamefont {Wang}}, \bibinfo {author}
  {\bibfnamefont {X.}~\bibnamefont {Liu}}, \bibinfo {author} {\bibfnamefont
  {A.}~\bibnamefont {Harzheim}}, \bibinfo {author} {\bibfnamefont
  {A.}~\bibnamefont {Lucas}}, \bibinfo {author} {\bibfnamefont
  {S.}~\bibnamefont {Sachdev}}, \bibinfo {author} {\bibfnamefont
  {P.}~\bibnamefont {Kim}}, \bibinfo {author} {\bibfnamefont {T.}~\bibnamefont
  {Taniguchi}}, \bibinfo {author} {\bibfnamefont {K.}~\bibnamefont {Watanabe}},
  \bibinfo {author} {\bibfnamefont {T.~A.}\ \bibnamefont {Ohki}},\ and\
  \bibinfo {author} {\bibfnamefont {K.~C.}\ \bibnamefont {Fong}},\ }\href
  {https://doi.org/10.1126/science.aad0343} {\bibfield  {journal} {\bibinfo
  {journal} {Science}\ }\textbf {\bibinfo {volume} {351}},\ \bibinfo {pages}
  {1058} (\bibinfo {year} {2016})},\ \Eprint
  {https://arxiv.org/abs/https://www.science.org/doi/pdf/10.1126/science.aad0343}
  {https://www.science.org/doi/pdf/10.1126/science.aad0343} \BibitemShut
  {NoStop}%
\bibitem [{\citenamefont {Bandurin}\ \emph {et~al.}(2018)\citenamefont
  {Bandurin}, \citenamefont {Shytov}, \citenamefont {Levitov}, \citenamefont
  {Kumar}, \citenamefont {Berdyugin}, \citenamefont {Ben~Shalom}, \citenamefont
  {Grigorieva}, \citenamefont {Geim},\ and\ \citenamefont
  {Falkovich}}]{Bandurin2018}%
  \BibitemOpen
  \bibfield  {author} {\bibinfo {author} {\bibfnamefont {D.~A.}\ \bibnamefont
  {Bandurin}}, \bibinfo {author} {\bibfnamefont {A.~V.}\ \bibnamefont
  {Shytov}}, \bibinfo {author} {\bibfnamefont {L.~S.}\ \bibnamefont {Levitov}},
  \bibinfo {author} {\bibfnamefont {R.~K.}\ \bibnamefont {Kumar}}, \bibinfo
  {author} {\bibfnamefont {A.~I.}\ \bibnamefont {Berdyugin}}, \bibinfo {author}
  {\bibfnamefont {M.}~\bibnamefont {Ben~Shalom}}, \bibinfo {author}
  {\bibfnamefont {I.~V.}\ \bibnamefont {Grigorieva}}, \bibinfo {author}
  {\bibfnamefont {A.~K.}\ \bibnamefont {Geim}},\ and\ \bibinfo {author}
  {\bibfnamefont {G.}~\bibnamefont {Falkovich}},\ }\href
  {https://doi.org/10.1038/s41467-018-07004-4} {\bibfield  {journal} {\bibinfo
  {journal} {Nature Communications}\ }\textbf {\bibinfo {volume} {9}},\
  \bibinfo {pages} {4533} (\bibinfo {year} {2018})}\BibitemShut {NoStop}%
\bibitem [{\citenamefont {Braem}\ \emph {et~al.}(2018)\citenamefont {Braem},
  \citenamefont {Pellegrino}, \citenamefont {Principi}, \citenamefont
  {R\"o\"osli}, \citenamefont {Gold}, \citenamefont {Hennel}, \citenamefont
  {Koski}, \citenamefont {Berl}, \citenamefont {Dietsche}, \citenamefont
  {Wegscheider}, \citenamefont {Polini}, \citenamefont {Ihn},\ and\
  \citenamefont {Ensslin}}]{Braem2018}%
  \BibitemOpen
  \bibfield  {author} {\bibinfo {author} {\bibfnamefont {B.~A.}\ \bibnamefont
  {Braem}}, \bibinfo {author} {\bibfnamefont {F.~M.~D.}\ \bibnamefont
  {Pellegrino}}, \bibinfo {author} {\bibfnamefont {A.}~\bibnamefont
  {Principi}}, \bibinfo {author} {\bibfnamefont {M.}~\bibnamefont
  {R\"o\"osli}}, \bibinfo {author} {\bibfnamefont {C.}~\bibnamefont {Gold}},
  \bibinfo {author} {\bibfnamefont {S.}~\bibnamefont {Hennel}}, \bibinfo
  {author} {\bibfnamefont {J.~V.}\ \bibnamefont {Koski}}, \bibinfo {author}
  {\bibfnamefont {M.}~\bibnamefont {Berl}}, \bibinfo {author} {\bibfnamefont
  {W.}~\bibnamefont {Dietsche}}, \bibinfo {author} {\bibfnamefont
  {W.}~\bibnamefont {Wegscheider}}, \bibinfo {author} {\bibfnamefont
  {M.}~\bibnamefont {Polini}}, \bibinfo {author} {\bibfnamefont
  {T.}~\bibnamefont {Ihn}},\ and\ \bibinfo {author} {\bibfnamefont
  {K.}~\bibnamefont {Ensslin}},\ }\href
  {https://doi.org/10.1103/PhysRevB.98.241304} {\bibfield  {journal} {\bibinfo
  {journal} {Phys. Rev. B}\ }\textbf {\bibinfo {volume} {98}},\ \bibinfo
  {pages} {241304} (\bibinfo {year} {2018})}\BibitemShut {NoStop}%
\bibitem [{\citenamefont {Sulpizio}\ \emph {et~al.}(2019)\citenamefont
  {Sulpizio}, \citenamefont {Ella}, \citenamefont {Rozen}, \citenamefont
  {Birkbeck}, \citenamefont {Perello}, \citenamefont {Dutta}, \citenamefont
  {Ben-Shalom}, \citenamefont {Taniguchi}, \citenamefont {Watanabe},
  \citenamefont {Holder}, \citenamefont {Queiroz}, \citenamefont {Principi},
  \citenamefont {Stern}, \citenamefont {Scaffidi}, \citenamefont {Geim},\ and\
  \citenamefont {Ilani}}]{Sulpizio2019}%
  \BibitemOpen
  \bibfield  {author} {\bibinfo {author} {\bibfnamefont {J.~A.}\ \bibnamefont
  {Sulpizio}}, \bibinfo {author} {\bibfnamefont {L.}~\bibnamefont {Ella}},
  \bibinfo {author} {\bibfnamefont {A.}~\bibnamefont {Rozen}}, \bibinfo
  {author} {\bibfnamefont {J.}~\bibnamefont {Birkbeck}}, \bibinfo {author}
  {\bibfnamefont {D.~J.}\ \bibnamefont {Perello}}, \bibinfo {author}
  {\bibfnamefont {D.}~\bibnamefont {Dutta}}, \bibinfo {author} {\bibfnamefont
  {M.}~\bibnamefont {Ben-Shalom}}, \bibinfo {author} {\bibfnamefont
  {T.}~\bibnamefont {Taniguchi}}, \bibinfo {author} {\bibfnamefont
  {K.}~\bibnamefont {Watanabe}}, \bibinfo {author} {\bibfnamefont
  {T.}~\bibnamefont {Holder}}, \bibinfo {author} {\bibfnamefont
  {R.}~\bibnamefont {Queiroz}}, \bibinfo {author} {\bibfnamefont
  {A.}~\bibnamefont {Principi}}, \bibinfo {author} {\bibfnamefont
  {A.}~\bibnamefont {Stern}}, \bibinfo {author} {\bibfnamefont
  {T.}~\bibnamefont {Scaffidi}}, \bibinfo {author} {\bibfnamefont {A.~K.}\
  \bibnamefont {Geim}},\ and\ \bibinfo {author} {\bibfnamefont
  {S.}~\bibnamefont {Ilani}},\ }\href
  {https://doi.org/10.1038/s41586-019-1788-9} {\bibfield  {journal} {\bibinfo
  {journal} {Nature}\ }\textbf {\bibinfo {volume} {576}},\ \bibinfo {pages}
  {75} (\bibinfo {year} {2019})}\BibitemShut {NoStop}%
\bibitem [{\citenamefont {Ku}\ \emph {et~al.}(2020)\citenamefont {Ku},
  \citenamefont {Zhou}, \citenamefont {Li}, \citenamefont {Shin}, \citenamefont
  {Shi}, \citenamefont {Burch}, \citenamefont {Anderson}, \citenamefont
  {Pierce}, \citenamefont {Xie}, \citenamefont {Hamo}, \citenamefont {Vool},
  \citenamefont {Zhang}, \citenamefont {Casola}, \citenamefont {Taniguchi},
  \citenamefont {Watanabe}, \citenamefont {Fogler}, \citenamefont {Kim},
  \citenamefont {Yacoby},\ and\ \citenamefont {Walsworth}}]{Ku2020}%
  \BibitemOpen
  \bibfield  {author} {\bibinfo {author} {\bibfnamefont {M.~J.~H.}\
  \bibnamefont {Ku}}, \bibinfo {author} {\bibfnamefont {T.~X.}\ \bibnamefont
  {Zhou}}, \bibinfo {author} {\bibfnamefont {Q.}~\bibnamefont {Li}}, \bibinfo
  {author} {\bibfnamefont {Y.~J.}\ \bibnamefont {Shin}}, \bibinfo {author}
  {\bibfnamefont {J.~K.}\ \bibnamefont {Shi}}, \bibinfo {author} {\bibfnamefont
  {C.}~\bibnamefont {Burch}}, \bibinfo {author} {\bibfnamefont {L.~E.}\
  \bibnamefont {Anderson}}, \bibinfo {author} {\bibfnamefont {A.~T.}\
  \bibnamefont {Pierce}}, \bibinfo {author} {\bibfnamefont {Y.}~\bibnamefont
  {Xie}}, \bibinfo {author} {\bibfnamefont {A.}~\bibnamefont {Hamo}}, \bibinfo
  {author} {\bibfnamefont {U.}~\bibnamefont {Vool}}, \bibinfo {author}
  {\bibfnamefont {H.}~\bibnamefont {Zhang}}, \bibinfo {author} {\bibfnamefont
  {F.}~\bibnamefont {Casola}}, \bibinfo {author} {\bibfnamefont
  {T.}~\bibnamefont {Taniguchi}}, \bibinfo {author} {\bibfnamefont
  {K.}~\bibnamefont {Watanabe}}, \bibinfo {author} {\bibfnamefont {M.~M.}\
  \bibnamefont {Fogler}}, \bibinfo {author} {\bibfnamefont {P.}~\bibnamefont
  {Kim}}, \bibinfo {author} {\bibfnamefont {A.}~\bibnamefont {Yacoby}},\ and\
  \bibinfo {author} {\bibfnamefont {R.~L.}\ \bibnamefont {Walsworth}},\ }\href
  {https://doi.org/10.1038/s41586-020-2507-2} {\bibfield  {journal} {\bibinfo
  {journal} {Nature}\ }\textbf {\bibinfo {volume} {583}},\ \bibinfo {pages}
  {537} (\bibinfo {year} {2020})}\BibitemShut {NoStop}%
\bibitem [{\citenamefont {Kashuba}(2008)}]{Kashuba2008}%
  \BibitemOpen
  \bibfield  {author} {\bibinfo {author} {\bibfnamefont {A.~B.}\ \bibnamefont
  {Kashuba}},\ }\href {https://doi.org/10.1103/PhysRevB.78.085415} {\bibfield
  {journal} {\bibinfo  {journal} {Phys. Rev. B}\ }\textbf {\bibinfo {volume}
  {78}},\ \bibinfo {pages} {085415} (\bibinfo {year} {2008})}\BibitemShut
  {NoStop}%
\bibitem [{\citenamefont {Fritz}\ \emph {et~al.}(2008)\citenamefont {Fritz},
  \citenamefont {Schmalian}, \citenamefont {M\"uller},\ and\ \citenamefont
  {Sachdev}}]{Fritz2008a}%
  \BibitemOpen
  \bibfield  {author} {\bibinfo {author} {\bibfnamefont {L.}~\bibnamefont
  {Fritz}}, \bibinfo {author} {\bibfnamefont {J.}~\bibnamefont {Schmalian}},
  \bibinfo {author} {\bibfnamefont {M.}~\bibnamefont {M\"uller}},\ and\
  \bibinfo {author} {\bibfnamefont {S.}~\bibnamefont {Sachdev}},\ }\href
  {https://doi.org/10.1103/PhysRevB.78.085416} {\bibfield  {journal} {\bibinfo
  {journal} {Phys. Rev. B}\ }\textbf {\bibinfo {volume} {78}},\ \bibinfo
  {pages} {085416} (\bibinfo {year} {2008})}\BibitemShut {NoStop}%
\bibitem [{\citenamefont {M\"uller}\ \emph {et~al.}(2008)\citenamefont
  {M\"uller}, \citenamefont {Fritz},\ and\ \citenamefont
  {Sachdev}}]{Fritz2008b}%
  \BibitemOpen
  \bibfield  {author} {\bibinfo {author} {\bibfnamefont {M.}~\bibnamefont
  {M\"uller}}, \bibinfo {author} {\bibfnamefont {L.}~\bibnamefont {Fritz}},\
  and\ \bibinfo {author} {\bibfnamefont {S.}~\bibnamefont {Sachdev}},\ }\href
  {https://doi.org/10.1103/PhysRevB.78.115406} {\bibfield  {journal} {\bibinfo
  {journal} {Phys. Rev. B}\ }\textbf {\bibinfo {volume} {78}},\ \bibinfo
  {pages} {115406} (\bibinfo {year} {2008})}\BibitemShut {NoStop}%
\bibitem [{\citenamefont {M\"uller}\ \emph {et~al.}(2009)\citenamefont
  {M\"uller}, \citenamefont {Schmalian},\ and\ \citenamefont
  {Fritz}}]{Fritz2009}%
  \BibitemOpen
  \bibfield  {author} {\bibinfo {author} {\bibfnamefont {M.}~\bibnamefont
  {M\"uller}}, \bibinfo {author} {\bibfnamefont {J.}~\bibnamefont
  {Schmalian}},\ and\ \bibinfo {author} {\bibfnamefont {L.}~\bibnamefont
  {Fritz}},\ }\href {https://doi.org/10.1103/PhysRevLett.103.025301} {\bibfield
   {journal} {\bibinfo  {journal} {Phys. Rev. Lett.}\ }\textbf {\bibinfo
  {volume} {103}},\ \bibinfo {pages} {025301} (\bibinfo {year}
  {2009})}\BibitemShut {NoStop}%
\bibitem [{\citenamefont {Narozhny}\ \emph {et~al.}(2015)\citenamefont
  {Narozhny}, \citenamefont {Gornyi}, \citenamefont {Titov}, \citenamefont
  {Sch\"utt},\ and\ \citenamefont {Mirlin}}]{Narozhny2015}%
  \BibitemOpen
  \bibfield  {author} {\bibinfo {author} {\bibfnamefont {B.~N.}\ \bibnamefont
  {Narozhny}}, \bibinfo {author} {\bibfnamefont {I.~V.}\ \bibnamefont
  {Gornyi}}, \bibinfo {author} {\bibfnamefont {M.}~\bibnamefont {Titov}},
  \bibinfo {author} {\bibfnamefont {M.}~\bibnamefont {Sch\"utt}},\ and\
  \bibinfo {author} {\bibfnamefont {A.~D.}\ \bibnamefont {Mirlin}},\ }\href
  {https://doi.org/10.1103/PhysRevB.91.035414} {\bibfield  {journal} {\bibinfo
  {journal} {Phys. Rev. B}\ }\textbf {\bibinfo {volume} {91}},\ \bibinfo
  {pages} {035414} (\bibinfo {year} {2015})}\BibitemShut {NoStop}%
\bibitem [{\citenamefont {Lucas}\ \emph {et~al.}(2016)\citenamefont {Lucas},
  \citenamefont {Crossno}, \citenamefont {Fong}, \citenamefont {Kim},\ and\
  \citenamefont {Sachdev}}]{Lucas2016}%
  \BibitemOpen
  \bibfield  {author} {\bibinfo {author} {\bibfnamefont {A.}~\bibnamefont
  {Lucas}}, \bibinfo {author} {\bibfnamefont {J.}~\bibnamefont {Crossno}},
  \bibinfo {author} {\bibfnamefont {K.~C.}\ \bibnamefont {Fong}}, \bibinfo
  {author} {\bibfnamefont {P.}~\bibnamefont {Kim}},\ and\ \bibinfo {author}
  {\bibfnamefont {S.}~\bibnamefont {Sachdev}},\ }\href
  {https://doi.org/10.1103/PhysRevB.93.075426} {\bibfield  {journal} {\bibinfo
  {journal} {Phys. Rev. B}\ }\textbf {\bibinfo {volume} {93}},\ \bibinfo
  {pages} {075426} (\bibinfo {year} {2016})}\BibitemShut {NoStop}%
\bibitem [{\citenamefont {Kiselev}\ and\ \citenamefont
  {Schmalian}(2019)}]{Kiselev2019}%
  \BibitemOpen
  \bibfield  {author} {\bibinfo {author} {\bibfnamefont {E.~I.}\ \bibnamefont
  {Kiselev}}\ and\ \bibinfo {author} {\bibfnamefont {J.}~\bibnamefont
  {Schmalian}},\ }\href {https://doi.org/10.1103/PhysRevLett.123.195302}
  {\bibfield  {journal} {\bibinfo  {journal} {Phys. Rev. Lett.}\ }\textbf
  {\bibinfo {volume} {123}},\ \bibinfo {pages} {195302} (\bibinfo {year}
  {2019})}\BibitemShut {NoStop}%
\bibitem [{\citenamefont {Kiselev}\ and\ \citenamefont
  {Schmalian}(2020)}]{kiselev2020}%
  \BibitemOpen
  \bibfield  {author} {\bibinfo {author} {\bibfnamefont {E.~I.}\ \bibnamefont
  {Kiselev}}\ and\ \bibinfo {author} {\bibfnamefont {J.}~\bibnamefont
  {Schmalian}},\ }\href {https://doi.org/10.1103/PhysRevB.102.245434}
  {\bibfield  {journal} {\bibinfo  {journal} {Phys. Rev. B}\ }\textbf {\bibinfo
  {volume} {102}},\ \bibinfo {pages} {245434} (\bibinfo {year}
  {2020})}\BibitemShut {NoStop}%
\bibitem [{\citenamefont {Afanasiev}\ \emph {et~al.}(2021)\citenamefont
  {Afanasiev}, \citenamefont {Alekseev}, \citenamefont {Greshnov},\ and\
  \citenamefont {Semina}}]{Alekseev2021}%
  \BibitemOpen
  \bibfield  {author} {\bibinfo {author} {\bibfnamefont {A.~N.}\ \bibnamefont
  {Afanasiev}}, \bibinfo {author} {\bibfnamefont {P.~S.}\ \bibnamefont
  {Alekseev}}, \bibinfo {author} {\bibfnamefont {A.~A.}\ \bibnamefont
  {Greshnov}},\ and\ \bibinfo {author} {\bibfnamefont {M.~A.}\ \bibnamefont
  {Semina}},\ }\href {https://doi.org/10.1103/PhysRevB.104.195415} {\bibfield
  {journal} {\bibinfo  {journal} {Phys. Rev. B}\ }\textbf {\bibinfo {volume}
  {104}},\ \bibinfo {pages} {195415} (\bibinfo {year} {2021})}\BibitemShut
  {NoStop}%
\bibitem [{\citenamefont {Xian}\ \emph {et~al.}(2023)\citenamefont {Xian},
  \citenamefont {Danz}, \citenamefont {Rodr\'{\i}guez~Fern\'andez},
  \citenamefont {Matthaiakakis}, \citenamefont {Tutschku}, \citenamefont
  {Klees}, \citenamefont {Erdmenger}, \citenamefont {Meyer},\ and\
  \citenamefont {Hankiewicz}}]{Hankiewicz2023}%
  \BibitemOpen
  \bibfield  {author} {\bibinfo {author} {\bibfnamefont {Z.-Y.}\ \bibnamefont
  {Xian}}, \bibinfo {author} {\bibfnamefont {S.}~\bibnamefont {Danz}}, \bibinfo
  {author} {\bibfnamefont {D.}~\bibnamefont {Rodr\'{\i}guez~Fern\'andez}},
  \bibinfo {author} {\bibfnamefont {I.}~\bibnamefont {Matthaiakakis}}, \bibinfo
  {author} {\bibfnamefont {C.}~\bibnamefont {Tutschku}}, \bibinfo {author}
  {\bibfnamefont {R.~L.}\ \bibnamefont {Klees}}, \bibinfo {author}
  {\bibfnamefont {J.}~\bibnamefont {Erdmenger}}, \bibinfo {author}
  {\bibfnamefont {R.}~\bibnamefont {Meyer}},\ and\ \bibinfo {author}
  {\bibfnamefont {E.~M.}\ \bibnamefont {Hankiewicz}},\ }\href
  {https://doi.org/10.1103/PhysRevB.107.L201403} {\bibfield  {journal}
  {\bibinfo  {journal} {Phys. Rev. B}\ }\textbf {\bibinfo {volume} {107}},\
  \bibinfo {pages} {L201403} (\bibinfo {year} {2023})}\BibitemShut {NoStop}%
\bibitem [{\citenamefont {Pines}\ and\ \citenamefont
  {Bohm}(1952)}]{PinesBohm1952}%
  \BibitemOpen
  \bibfield  {author} {\bibinfo {author} {\bibfnamefont {D.}~\bibnamefont
  {Pines}}\ and\ \bibinfo {author} {\bibfnamefont {D.}~\bibnamefont {Bohm}},\
  }\href {https://doi.org/10.1103/PhysRev.85.338} {\bibfield  {journal}
  {\bibinfo  {journal} {Phys. Rev.}\ }\textbf {\bibinfo {volume} {85}},\
  \bibinfo {pages} {338} (\bibinfo {year} {1952})}\BibitemShut {NoStop}%
\bibitem [{\citenamefont {Bohm}\ and\ \citenamefont
  {Pines}(1953)}]{BohmPine1953}%
  \BibitemOpen
  \bibfield  {author} {\bibinfo {author} {\bibfnamefont {D.}~\bibnamefont
  {Bohm}}\ and\ \bibinfo {author} {\bibfnamefont {D.}~\bibnamefont {Pines}},\
  }\href {https://doi.org/10.1103/PhysRev.92.609} {\bibfield  {journal}
  {\bibinfo  {journal} {Phys. Rev.}\ }\textbf {\bibinfo {volume} {92}},\
  \bibinfo {pages} {609} (\bibinfo {year} {1953})}\BibitemShut {NoStop}%
\bibitem [{\citenamefont {Vlasov}(1968)}]{Vlasov1968}%
  \BibitemOpen
  \bibfield  {author} {\bibinfo {author} {\bibfnamefont {A.~A.}\ \bibnamefont
  {Vlasov}},\ }\href {https://doi.org/10.1070/pu1968v010n06abeh003709}
  {\bibfield  {journal} {\bibinfo  {journal} {Soviet Physics Uspekhi}\ }\textbf
  {\bibinfo {volume} {10}},\ \bibinfo {pages} {721} (\bibinfo {year}
  {1968})}\BibitemShut {NoStop}%
\bibitem [{\citenamefont {Pines}(1953)}]{Pines1953}%
  \BibitemOpen
  \bibfield  {author} {\bibinfo {author} {\bibfnamefont {D.}~\bibnamefont
  {Pines}},\ }\href {https://doi.org/10.1103/PhysRev.92.626} {\bibfield
  {journal} {\bibinfo  {journal} {Phys. Rev.}\ }\textbf {\bibinfo {volume}
  {92}},\ \bibinfo {pages} {626} (\bibinfo {year} {1953})}\BibitemShut
  {NoStop}%
\bibitem [{\citenamefont {Stern}(1967)}]{Stern1967}%
  \BibitemOpen
  \bibfield  {author} {\bibinfo {author} {\bibfnamefont {F.}~\bibnamefont
  {Stern}},\ }\href@noop {} {\bibfield  {journal} {\bibinfo  {journal} {Phys.
  Rev. Lett.}\ }\textbf {\bibinfo {volume} {18}},\ \bibinfo {pages} {546}
  (\bibinfo {year} {1967})}\BibitemShut {NoStop}%
\bibitem [{\citenamefont {Wunsch}\ \emph {et~al.}(2006)\citenamefont {Wunsch},
  \citenamefont {Stauber}, \citenamefont {Sols},\ and\ \citenamefont
  {Guinea}}]{Wunsch2006}%
  \BibitemOpen
  \bibfield  {author} {\bibinfo {author} {\bibfnamefont {B.}~\bibnamefont
  {Wunsch}}, \bibinfo {author} {\bibfnamefont {T.}~\bibnamefont {Stauber}},
  \bibinfo {author} {\bibfnamefont {F.}~\bibnamefont {Sols}},\ and\ \bibinfo
  {author} {\bibfnamefont {F.}~\bibnamefont {Guinea}},\ }\href
  {https://doi.org/10.1088/1367-2630/8/12/318} {\bibfield  {journal} {\bibinfo
  {journal} {New Journal of Physics}\ }\textbf {\bibinfo {volume} {8}},\
  \bibinfo {pages} {318} (\bibinfo {year} {2006})}\BibitemShut {NoStop}%
\bibitem [{\citenamefont {Das~Sarma}\ and\ \citenamefont
  {Li}(2013)}]{DasSarma2013}%
  \BibitemOpen
  \bibfield  {author} {\bibinfo {author} {\bibfnamefont {S.}~\bibnamefont
  {Das~Sarma}}\ and\ \bibinfo {author} {\bibfnamefont {Q.}~\bibnamefont {Li}},\
  }\href {https://doi.org/10.1103/PhysRevB.87.235418} {\bibfield  {journal}
  {\bibinfo  {journal} {Phys. Rev. B}\ }\textbf {\bibinfo {volume} {87}},\
  \bibinfo {pages} {235418} (\bibinfo {year} {2013})}\BibitemShut {NoStop}%
\bibitem [{\citenamefont {Takada}(1978)}]{Takada1978}%
  \BibitemOpen
  \bibfield  {author} {\bibinfo {author} {\bibfnamefont {Y.}~\bibnamefont
  {Takada}},\ }\href {https://doi.org/10.1143/JPSJ.45.786} {\bibfield
  {journal} {\bibinfo  {journal} {Journal of the Physical Society of Japan}\
  }\textbf {\bibinfo {volume} {45}},\ \bibinfo {pages} {786} (\bibinfo {year}
  {1978})},\ \Eprint
  {https://arxiv.org/abs/https://doi.org/10.1143/JPSJ.45.786}
  {https://doi.org/10.1143/JPSJ.45.786} \BibitemShut {NoStop}%
\bibitem [{\citenamefont {Ruhman}\ and\ \citenamefont {Lee}(2017)}]{Lee2017}%
  \BibitemOpen
  \bibfield  {author} {\bibinfo {author} {\bibfnamefont {J.}~\bibnamefont
  {Ruhman}}\ and\ \bibinfo {author} {\bibfnamefont {P.~A.}\ \bibnamefont
  {Lee}},\ }\href {https://doi.org/10.1103/PhysRevB.96.235107} {\bibfield
  {journal} {\bibinfo  {journal} {Phys. Rev. B}\ }\textbf {\bibinfo {volume}
  {96}},\ \bibinfo {pages} {235107} (\bibinfo {year} {2017})}\BibitemShut
  {NoStop}%
\bibitem [{\citenamefont {Wyld}\ and\ \citenamefont
  {Pines}(1962)}]{WyldPines1962}%
  \BibitemOpen
  \bibfield  {author} {\bibinfo {author} {\bibfnamefont {H.~W.}\ \bibnamefont
  {Wyld}}\ and\ \bibinfo {author} {\bibfnamefont {D.}~\bibnamefont {Pines}},\
  }\href {https://doi.org/10.1103/PhysRev.127.1851} {\bibfield  {journal}
  {\bibinfo  {journal} {Phys. Rev.}\ }\textbf {\bibinfo {volume} {127}},\
  \bibinfo {pages} {1851} (\bibinfo {year} {1962})}\BibitemShut {NoStop}%
\bibitem [{\citenamefont {Rana}\ \emph {et~al.}(2011)\citenamefont {Rana},
  \citenamefont {Strait}, \citenamefont {Wang},\ and\ \citenamefont
  {Manolatou}}]{Rana2011}%
  \BibitemOpen
  \bibfield  {author} {\bibinfo {author} {\bibfnamefont {F.}~\bibnamefont
  {Rana}}, \bibinfo {author} {\bibfnamefont {J.~H.}\ \bibnamefont {Strait}},
  \bibinfo {author} {\bibfnamefont {H.}~\bibnamefont {Wang}},\ and\ \bibinfo
  {author} {\bibfnamefont {C.}~\bibnamefont {Manolatou}},\ }\href
  {https://doi.org/10.1103/PhysRevB.84.045437} {\bibfield  {journal} {\bibinfo
  {journal} {Phys. Rev. B}\ }\textbf {\bibinfo {volume} {84}},\ \bibinfo
  {pages} {045437} (\bibinfo {year} {2011})}\BibitemShut {NoStop}%
\bibitem [{\citenamefont {Klug}\ \emph {et~al.}(2018)\citenamefont {Klug},
  \citenamefont {Scheurer},\ and\ \citenamefont {Schmalian}}]{Schmalian2018}%
  \BibitemOpen
  \bibfield  {author} {\bibinfo {author} {\bibfnamefont {M.~J.}\ \bibnamefont
  {Klug}}, \bibinfo {author} {\bibfnamefont {M.~S.}\ \bibnamefont {Scheurer}},\
  and\ \bibinfo {author} {\bibfnamefont {J.}~\bibnamefont {Schmalian}},\ }\href
  {https://doi.org/10.1103/PhysRevB.98.045102} {\bibfield  {journal} {\bibinfo
  {journal} {Phys. Rev. B}\ }\textbf {\bibinfo {volume} {98}},\ \bibinfo
  {pages} {045102} (\bibinfo {year} {2018})}\BibitemShut {NoStop}%
\bibitem [{\citenamefont {Pongsangangan}\ \emph {et~al.}(2020)\citenamefont
  {Pongsangangan}, \citenamefont {Grubinskas},\ and\ \citenamefont
  {Fritz}}]{Kitinan2022a}%
  \BibitemOpen
  \bibfield  {author} {\bibinfo {author} {\bibfnamefont {K.}~\bibnamefont
  {Pongsangangan}}, \bibinfo {author} {\bibfnamefont {S.}~\bibnamefont
  {Grubinskas}},\ and\ \bibinfo {author} {\bibfnamefont {L.}~\bibnamefont
  {Fritz}},\ }\href {https://arxiv.org/abs/2005.12790} {\bibinfo {title}
  {Thermo-electric response in two-dimensional dirac systems: the role of
  particle-hole pairs}} (\bibinfo {year} {2020})\BibitemShut {NoStop}%
\bibitem [{\citenamefont {Pongsangangan}\ \emph
  {et~al.}(2022{\natexlab{a}})\citenamefont {Pongsangangan}, \citenamefont
  {Ludwig}, \citenamefont {Stoof},\ and\ \citenamefont {Fritz}}]{Kitinan2022b}%
  \BibitemOpen
  \bibfield  {author} {\bibinfo {author} {\bibfnamefont {K.}~\bibnamefont
  {Pongsangangan}}, \bibinfo {author} {\bibfnamefont {T.}~\bibnamefont
  {Ludwig}}, \bibinfo {author} {\bibfnamefont {H.~T.~C.}\ \bibnamefont
  {Stoof}},\ and\ \bibinfo {author} {\bibfnamefont {L.}~\bibnamefont {Fritz}},\
  }\href@noop {} {\bibinfo {title} {Hydrodynamics of charged two-dimensional
  dirac systems $\text{I}$: thermo-electric transport}} (\bibinfo {year}
  {2022}{\natexlab{a}})\BibitemShut {NoStop}%
\bibitem [{\citenamefont {Pongsangangan}\ \emph
  {et~al.}(2022{\natexlab{b}})\citenamefont {Pongsangangan}, \citenamefont
  {Ludwig}, \citenamefont {Stoof},\ and\ \citenamefont {Fritz}}]{Kitinan2022c}%
  \BibitemOpen
  \bibfield  {author} {\bibinfo {author} {\bibfnamefont {K.}~\bibnamefont
  {Pongsangangan}}, \bibinfo {author} {\bibfnamefont {T.}~\bibnamefont
  {Ludwig}}, \bibinfo {author} {\bibfnamefont {H.~T.~C.}\ \bibnamefont
  {Stoof}},\ and\ \bibinfo {author} {\bibfnamefont {L.}~\bibnamefont {Fritz}},\
  }\href {https://arxiv.org/abs/2206.09694} {\bibinfo {title} {Hydrodynamics of
  charged two-dimensional dirac systems $\text{II}$: the role of collective
  modes}} (\bibinfo {year} {2022}{\natexlab{b}})\BibitemShut {NoStop}%
\bibitem [{\citenamefont {Link}\ \emph {et~al.}(2018)\citenamefont {Link},
  \citenamefont {Sheehy}, \citenamefont {Narozhny},\ and\ \citenamefont
  {Schmalian}}]{Link2018}%
  \BibitemOpen
  \bibfield  {author} {\bibinfo {author} {\bibfnamefont {J.~M.}\ \bibnamefont
  {Link}}, \bibinfo {author} {\bibfnamefont {D.~E.}\ \bibnamefont {Sheehy}},
  \bibinfo {author} {\bibfnamefont {B.~N.}\ \bibnamefont {Narozhny}},\ and\
  \bibinfo {author} {\bibfnamefont {J.}~\bibnamefont {Schmalian}},\ }\href
  {https://doi.org/10.1103/PhysRevB.98.195103} {\bibfield  {journal} {\bibinfo
  {journal} {Phys. Rev. B}\ }\textbf {\bibinfo {volume} {98}},\ \bibinfo
  {pages} {195103} (\bibinfo {year} {2018})}\BibitemShut {NoStop}%
\bibitem [{\citenamefont {Pines}\ and\ \citenamefont
  {Schrieffer}(1962)}]{PinesSchrieffer1962}%
  \BibitemOpen
  \bibfield  {author} {\bibinfo {author} {\bibfnamefont {D.}~\bibnamefont
  {Pines}}\ and\ \bibinfo {author} {\bibfnamefont {J.~R.}\ \bibnamefont
  {Schrieffer}},\ }\href {https://doi.org/10.1103/PhysRev.125.804} {\bibfield
  {journal} {\bibinfo  {journal} {Phys. Rev.}\ }\textbf {\bibinfo {volume}
  {125}},\ \bibinfo {pages} {804} (\bibinfo {year} {1962})}\BibitemShut
  {NoStop}%
\bibitem [{\citenamefont {Castro~Neto}\ \emph {et~al.}(2009)\citenamefont
  {Castro~Neto}, \citenamefont {Guinea}, \citenamefont {Peres}, \citenamefont
  {Novoselov},\ and\ \citenamefont {Geim}}]{CastroNeto2009}%
  \BibitemOpen
  \bibfield  {author} {\bibinfo {author} {\bibfnamefont {A.~H.}\ \bibnamefont
  {Castro~Neto}}, \bibinfo {author} {\bibfnamefont {F.}~\bibnamefont {Guinea}},
  \bibinfo {author} {\bibfnamefont {N.~M.~R.}\ \bibnamefont {Peres}}, \bibinfo
  {author} {\bibfnamefont {K.~S.}\ \bibnamefont {Novoselov}},\ and\ \bibinfo
  {author} {\bibfnamefont {A.~K.}\ \bibnamefont {Geim}},\ }\href
  {https://doi.org/10.1103/RevModPhys.81.109} {\bibfield  {journal} {\bibinfo
  {journal} {Rev. Mod. Phys.}\ }\textbf {\bibinfo {volume} {81}},\ \bibinfo
  {pages} {109} (\bibinfo {year} {2009})}\BibitemShut {NoStop}%
\bibitem [{\citenamefont {Kotov}\ \emph {et~al.}(2012)\citenamefont {Kotov},
  \citenamefont {Uchoa}, \citenamefont {Pereira}, \citenamefont {Guinea},\ and\
  \citenamefont {Castro~Neto}}]{Kotov2012}%
  \BibitemOpen
  \bibfield  {author} {\bibinfo {author} {\bibfnamefont {V.~N.}\ \bibnamefont
  {Kotov}}, \bibinfo {author} {\bibfnamefont {B.}~\bibnamefont {Uchoa}},
  \bibinfo {author} {\bibfnamefont {V.~M.}\ \bibnamefont {Pereira}}, \bibinfo
  {author} {\bibfnamefont {F.}~\bibnamefont {Guinea}},\ and\ \bibinfo {author}
  {\bibfnamefont {A.~H.}\ \bibnamefont {Castro~Neto}},\ }\href
  {https://doi.org/10.1103/RevModPhys.84.1067} {\bibfield  {journal} {\bibinfo
  {journal} {Rev. Mod. Phys.}\ }\textbf {\bibinfo {volume} {84}},\ \bibinfo
  {pages} {1067} (\bibinfo {year} {2012})}\BibitemShut {NoStop}%
\bibitem [{\citenamefont {Chapman}\ and\ \citenamefont
  {Cowling}(1954)}]{Chapman1954TheGases}%
  \BibitemOpen
  \bibfield  {author} {\bibinfo {author} {\bibfnamefont {S.}~\bibnamefont
  {Chapman}}\ and\ \bibinfo {author} {\bibfnamefont {T.~G.}\ \bibnamefont
  {Cowling}},\ }\href@noop {} {\emph {\bibinfo {title} {{The Mathematical
  Theory of Non-Uniform Gases. An Account of the Kinetic Theory of Viscosity,
  Thermal Conduction, and Diffusion in Gases}}}}\ (\bibinfo  {publisher}
  {Cambridge Universiry Press},\ \bibinfo {year} {1954})\BibitemShut {NoStop}%
\bibitem [{\citenamefont {Bhatnagar}\ \emph {et~al.}(1954)\citenamefont
  {Bhatnagar}, \citenamefont {Gross},\ and\ \citenamefont
  {Krook}}]{PhysRev.94.511}%
  \BibitemOpen
  \bibfield  {author} {\bibinfo {author} {\bibfnamefont {P.~L.}\ \bibnamefont
  {Bhatnagar}}, \bibinfo {author} {\bibfnamefont {E.~P.}\ \bibnamefont
  {Gross}},\ and\ \bibinfo {author} {\bibfnamefont {M.}~\bibnamefont {Krook}},\
  }\href {https://doi.org/10.1103/PhysRev.94.511} {\bibfield  {journal}
  {\bibinfo  {journal} {Phys. Rev.}\ }\textbf {\bibinfo {volume} {94}},\
  \bibinfo {pages} {511} (\bibinfo {year} {1954})}\BibitemShut {NoStop}%
\bibitem [{\citenamefont {Gross}\ and\ \citenamefont
  {Krook}(1956)}]{Gross1956}%
  \BibitemOpen
  \bibfield  {author} {\bibinfo {author} {\bibfnamefont {E.~P.}\ \bibnamefont
  {Gross}}\ and\ \bibinfo {author} {\bibfnamefont {M.}~\bibnamefont {Krook}},\
  }\href {https://doi.org/10.1103/PhysRev.102.593} {\bibfield  {journal}
  {\bibinfo  {journal} {Physical Review}\ }\textbf {\bibinfo {volume} {102}},\
  \bibinfo {pages} {593} (\bibinfo {year} {1956})}\BibitemShut {NoStop}%
\bibitem [{\citenamefont {Sch\"utt}\ \emph {et~al.}(2011)\citenamefont
  {Sch\"utt}, \citenamefont {Ostrovsky}, \citenamefont {Gornyi},\ and\
  \citenamefont {Mirlin}}]{schuett2011}%
  \BibitemOpen
  \bibfield  {author} {\bibinfo {author} {\bibfnamefont {M.}~\bibnamefont
  {Sch\"utt}}, \bibinfo {author} {\bibfnamefont {P.~M.}\ \bibnamefont
  {Ostrovsky}}, \bibinfo {author} {\bibfnamefont {I.~V.}\ \bibnamefont
  {Gornyi}},\ and\ \bibinfo {author} {\bibfnamefont {A.~D.}\ \bibnamefont
  {Mirlin}},\ }\href {https://doi.org/10.1103/PhysRevB.83.155441} {\bibfield
  {journal} {\bibinfo  {journal} {Phys. Rev. B}\ }\textbf {\bibinfo {volume}
  {83}},\ \bibinfo {pages} {155441} (\bibinfo {year} {2011})}\BibitemShut
  {NoStop}%
\bibitem [{\citenamefont {Kozii}\ and\ \citenamefont {Fu}(2018)}]{Fu2018}%
  \BibitemOpen
  \bibfield  {author} {\bibinfo {author} {\bibfnamefont {V.}~\bibnamefont
  {Kozii}}\ and\ \bibinfo {author} {\bibfnamefont {L.}~\bibnamefont {Fu}},\
  }\href {https://doi.org/10.1103/PhysRevB.98.041109} {\bibfield  {journal}
  {\bibinfo  {journal} {Phys. Rev. B}\ }\textbf {\bibinfo {volume} {98}},\
  \bibinfo {pages} {041109} (\bibinfo {year} {2018})}\BibitemShut {NoStop}%
\bibitem [{\citenamefont {Gabbana}\ \emph {et~al.}(2020)\citenamefont
  {Gabbana}, \citenamefont {Simeoni}, \citenamefont {Succi},\ and\
  \citenamefont {Tripiccione}}]{Gabbana2020RelativisticApplications}%
  \BibitemOpen
  \bibfield  {author} {\bibinfo {author} {\bibfnamefont {A.}~\bibnamefont
  {Gabbana}}, \bibinfo {author} {\bibfnamefont {D.}~\bibnamefont {Simeoni}},
  \bibinfo {author} {\bibfnamefont {S.}~\bibnamefont {Succi}},\ and\ \bibinfo
  {author} {\bibfnamefont {R.}~\bibnamefont {Tripiccione}},\ }\href
  {https://doi.org/10.1016/j.physrep.2020.03.004} {\bibfield  {journal}
  {\bibinfo  {journal} {Physics Reports}\ }\textbf {\bibinfo {volume} {863}},\
  \bibinfo {pages} {1} (\bibinfo {year} {2020})}\BibitemShut {NoStop}%
\bibitem [{\citenamefont {Chen}\ and\ \citenamefont
  {Zhu}(2022)}]{Chen2022ViscosityElectrons}%
  \BibitemOpen
  \bibfield  {author} {\bibinfo {author} {\bibfnamefont {W.}~\bibnamefont
  {Chen}}\ and\ \bibinfo {author} {\bibfnamefont {W.}~\bibnamefont {Zhu}},\
  }\href {https://doi.org/10.1103/PhysRevB.106.014205} {\bibfield  {journal}
  {\bibinfo  {journal} {Physical Review B}\ }\textbf {\bibinfo {volume}
  {106}},\ \bibinfo {pages} {014205} (\bibinfo {year} {2022})}\BibitemShut
  {NoStop}%
\bibitem [{Note1()}]{Note1}%
  \BibitemOpen
  \bibinfo {note} {As we have seen the CE method shows that the bulk viscosity
  of the fermionic sector should be identically null and that the for the
  plasmon sector it is $\sim 1/10$ of the shear counterpart.}\BibitemShut
  {Stop}%
\bibitem [{\citenamefont {Ziman}(2000)}]{Ziman2000}%
  \BibitemOpen
  \bibfield  {author} {\bibinfo {author} {\bibfnamefont {J.}~\bibnamefont
  {Ziman}},\ }\href@noop {} {\emph {\bibinfo {title} {Electrons and phonons}}}\
  (\bibinfo  {publisher} {Oxford University Press},\ \bibinfo {year}
  {2000})\BibitemShut {NoStop}%
\bibitem [{\citenamefont {Liao}\ and\ \citenamefont
  {Galitski}(2020)}]{Galitski2020}%
  \BibitemOpen
  \bibfield  {author} {\bibinfo {author} {\bibfnamefont {Y.}~\bibnamefont
  {Liao}}\ and\ \bibinfo {author} {\bibfnamefont {V.}~\bibnamefont
  {Galitski}},\ }\href {https://doi.org/10.1103/PhysRevB.101.195106} {\bibfield
   {journal} {\bibinfo  {journal} {Phys. Rev. B}\ }\textbf {\bibinfo {volume}
  {101}},\ \bibinfo {pages} {195106} (\bibinfo {year} {2020})}\BibitemShut
  {NoStop}%
\bibitem [{\citenamefont {Zverevich}\ and\ \citenamefont
  {Levchenko}(2023)}]{zverevich2023transport}%
  \BibitemOpen
  \bibfield  {author} {\bibinfo {author} {\bibfnamefont {D.}~\bibnamefont
  {Zverevich}}\ and\ \bibinfo {author} {\bibfnamefont {A.}~\bibnamefont
  {Levchenko}},\ }\href@noop {} {\bibinfo {title} {Transport signatures of
  plasmon fluctuations in electron hydrodynamics}} (\bibinfo {year} {2023}),\
  \Eprint {https://arxiv.org/abs/2306.13534} {arXiv:2306.13534
  [cond-mat.mes-hall]} \BibitemShut {NoStop}%
\bibitem [{\citenamefont {Levchenko}\ and\ \citenamefont
  {Schmalian}(2020)}]{Levchenko2020}%
  \BibitemOpen
  \bibfield  {author} {\bibinfo {author} {\bibfnamefont {A.}~\bibnamefont
  {Levchenko}}\ and\ \bibinfo {author} {\bibfnamefont {J.}~\bibnamefont
  {Schmalian}},\ }\href
  {https://doi.org/https://doi.org/10.1016/j.aop.2020.168218} {\bibfield
  {journal} {\bibinfo  {journal} {Annals of Physics}\ }\textbf {\bibinfo
  {volume} {419}},\ \bibinfo {pages} {168218} (\bibinfo {year}
  {2020})}\BibitemShut {NoStop}%
\bibitem [{\citenamefont {Baym}\ and\ \citenamefont
  {Kadanoff}(1961)}]{KadanoffBaym1961}%
  \BibitemOpen
  \bibfield  {author} {\bibinfo {author} {\bibfnamefont {G.}~\bibnamefont
  {Baym}}\ and\ \bibinfo {author} {\bibfnamefont {L.~P.}\ \bibnamefont
  {Kadanoff}},\ }\href {https://doi.org/10.1103/PhysRev.124.287} {\bibfield
  {journal} {\bibinfo  {journal} {Phys. Rev.}\ }\textbf {\bibinfo {volume}
  {124}},\ \bibinfo {pages} {287} (\bibinfo {year} {1961})}\BibitemShut
  {NoStop}%
\bibitem [{\citenamefont {Narozhny}(2023)}]{Narozhny2023}%
  \BibitemOpen
  \bibfield  {author} {\bibinfo {author} {\bibfnamefont {B.~N.}\ \bibnamefont
  {Narozhny}},\ }\href
  {https://doi.org/https://doi.org/10.1016/j.aop.2023.169341} {\bibfield
  {journal} {\bibinfo  {journal} {Annals of Physics}\ }\textbf {\bibinfo
  {volume} {454}},\ \bibinfo {pages} {169341} (\bibinfo {year}
  {2023})}\BibitemShut {NoStop}%
\bibitem [{\citenamefont {Stoof}\ \emph {et~al.}(2009)\citenamefont {Stoof},
  \citenamefont {Gubbels},\ and\ \citenamefont {Dickerscheid}}]{Stoof2009}%
  \BibitemOpen
  \bibfield  {author} {\bibinfo {author} {\bibfnamefont {H.}~\bibnamefont
  {Stoof}}, \bibinfo {author} {\bibfnamefont {K.}~\bibnamefont {Gubbels}},\
  and\ \bibinfo {author} {\bibfnamefont {D.}~\bibnamefont {Dickerscheid}},\
  }\href@noop {} {\emph {\bibinfo {title} {Ultracold Quantum Fields}}}\
  (\bibinfo  {publisher} {Springer Science+Business Media B.V},\ \bibinfo
  {year} {2009})\BibitemShut {NoStop}%
\end{thebibliography}%

\appendix

\section{Fermi velocity renormalization}
\label{app:vfrenormalization}

Self-energy contributions are also present in Boltzmann equations. In this section, we show, however, that they can be absorbed into the Fermi velocity renormalization and, therefore, have a negligible impact on the global behavior of the viscosity. 
To this end, we consider the electron/hole self-energy's real part. As shown in Ref.~\cite{Kitinan2022c}, it comprises two terms reading as
\begin{equation}
\label{SE}
\mathrm{Re}\{\sigma(\vec{k},\epsilon_{\lambda,\vec{k}})\} = \sigma^{(1)}(\vec{k},\epsilon_{\lambda,\vec{k}}) + \sigma^{(2)}(\vec{k},\epsilon_{\lambda,\vec{k}}).
\end{equation}
The first term depends on the plasmonic distribution and contains divergent parts. The second term, on the other hand, depends only on the fermionic distribution function, and it is entirely regular.
As, in this section, we will focus our attention on the renormalization of the Fermi velocity, we will consider only $\sigma^{(1)}$. It can be cast into the following form~\cite{Kitinan2022c}:
\begin{multline}
        \label{SP}
		\sigma^{(1)}(\vec{k},\epsilon_{\lambda,\vec{k}})=\sum_{\lambda'}\int\frac{d\vec{q}}{(2\pi)^2}\frac{\pi\alpha\omega_\vec{q}}{2q}( 1 + 2b_\vec{q})\times \\
		\times\left[\frac{\mathcal{F}_{\lambda\lambda'}(\vec{k}+\vec{q},\vec{k})}{\epsilon_{\lambda,\vec{k}}+\omega_\vec{q}-\epsilon_{\lambda',\vec{k}+\vec{q}}} +  \frac{\mathcal{F}_{\lambda\lambda'}(\vec{k}-\vec{q},\vec{k})}{\epsilon_{\lambda,\vec{k}}-\omega_\vec{q}-\epsilon_{\lambda',\vec{k}-\vec{q}}} \right],
\end{multline}
where
$\mathcal{F}_{\lambda,\lambda'}$ are the form factors~\cite{Kitinan2022a}.
Here $\vec{k}$ is the momentum of the ingoing fermion and $\vec{q}$ is the momentum of the plasmon.
The electron-hole-plasmon system provides a natural IR cut-off $q_c$, then only the UV one needs to be introduced.
For small values of $\vec{k}$, we can use it as UV regulator. We expand Eq.(\ref{SP}) and extract the divergency as
	\begin{equation}
		\label{SPDiv}
		\sigma_{\rm div}^{(1)}(\vec{k},\epsilon_{\lambda,\vec{k}})= - \gamma \lambda k\int_{k}^{q_c}\frac{dq}{q} = -\gamma \lambda k \log\frac{q_c}{k},
	\end{equation}
with
\begin{equation}
    \label{Rgamma}
    \gamma^{-1} = \frac{N}{2}\log\left(2+2\cosh\frac{\mu}{T}\right).
\end{equation}
This term is directly responsible for Fermi velocity renormalization. It scales as $\frac{1}{N}\log \frac{1}{N\alpha}$. 

In addition to that, one may add regular terms which renormalize the kinetic energy as $\lambda k \rightarrow \lambda k  + \mathrm{Re}\{\sigma(\vec{k},\epsilon_{\lambda,\vec{k}}) \}$.



\section{Collision integrals}\label{app:collisions}

This appendix is devoted to the detailed description of the collision integrals for the electron-plasmon interaction.
 The collision integral for the plasmon equation is given in Eq.\eqref{eq:colbos}. There, the first term in the square bracket describes a scattering of a plasmon of momentum $\vec{q}$ with a fermion of momentum $\vec{k}$ and energy $\epsilon_{\lambda,\vec{k}}$ which, in turn, produces another fermion of momentum $\vec{k}+\vec{q}$ and energy $\epsilon_{\lambda',\vec{k}+\vec{q}}$; The second term describes the inverse process.
\begin{widetext}
	\begin{equation}
	\label{eq:colbos}
		\mathcal{I}^b[f,b]=-N\sum_{\lambda,\lambda'}\int \frac{d\vec{k}}{(2\pi)^2}  \mathcal{M}_{\lambda\lambda'}^{\vec{k}+\vec{q},\vec{k}}\delta(\omega_{\vec{q}}+\epsilon_{\lambda,\vec{k}}-\epsilon_{\lambda',\vec{k}+\vec{q}} )\left[f_{\lambda,\vec{k}}\left(1-f_{\lambda',\vec{k}+\vec{q}}\right)b_\vec{q}  - \left(1-f_{\lambda,\vec{k}}\right)f_{\lambda',\vec{k}+\vec{q}}\left(1+b_\vec{q}\right) \right],
	\end{equation}
	\begin{multline}
	\label{eq:colferm}
		\mathcal{I}^f_\lambda[f,b]=-\sum_{\lambda'}\int \frac{d\vec{q}}{(2\pi)^2}  \mathcal{M}_{\lambda\lambda'}^{\vec{k}+\vec{q},\vec{k}}\delta(\omega_{\vec{q}}+\epsilon_{\lambda,\vec{k}}-\epsilon_{\lambda',\vec{k}+\vec{q}} )\left[f_{\lambda,\vec{k}}\left(1-f_{\lambda',\vec{k}+\vec{q}}\right)b_\vec{q}  - \left(1-f_{\lambda,\vec{k}}\right)f_{\lambda',\vec{k}+\vec{q}}\left(1+b_\vec{q}\right) \right]\\
		-\sum_{\lambda'}\int \frac{d\vec{q}}{(2\pi)^2}  \mathcal{M}_{\lambda\lambda'}^{\vec{k}-\vec{q},\vec{k}}\delta(-\omega_{\vec{q}}+\epsilon_{\lambda,\vec{k}}-\epsilon_{\lambda',\vec{k}-\vec{q}} )\left[f_{\lambda,\vec{k}}\left(1-f_{\lambda',\vec{k}-\vec{q}}\right)\left(1+b_\vec{q}\right)  - \left(1-f_{\lambda,\vec{k}}\right)f_{\lambda',\vec{k}-\vec{q}}b_\vec{q} \right].
\end{multline}
\end{widetext}
%
%
The collision integral of the fermion equation is given by Eq.\eqref{eq:colferm} where
the first term describes a scattering of an electron from the momentum state $\vec{k}$ into another momentum state $\vec{k}+\vec{q}$ by absorbing a plasmon of momentum $\vec{q}$ and vice versa. The second term describes an emission of a plasmon of momentum $\vec{q}$ from an electron of momentum $\vec{k}$ and as a result the electron scatters into the momentum state $\vec{k}-\vec{q}$. In both integrals the cross-section is given by 
\begin{equation}
	\mathcal{M}_{\lambda\lambda'}^{\vec{k}+\vec{q},\vec{k}}=\frac{2\alpha \pi^2 \omega_{\vec{q}}}{q}\frac{1}{2}\left(1+\lambda\lambda'\cos\left(\theta_{\vec{k}+\vec{q}}-\theta_{\vec{k}}\right)\right),
\end{equation}
with $\theta_{\vec{k}} = \tan^{-1}(k_y/k_x)$ defining the direction of the momentum $\mathbf{k}$.
\begin{figure}
    \centering
    \includegraphics[scale=.5]{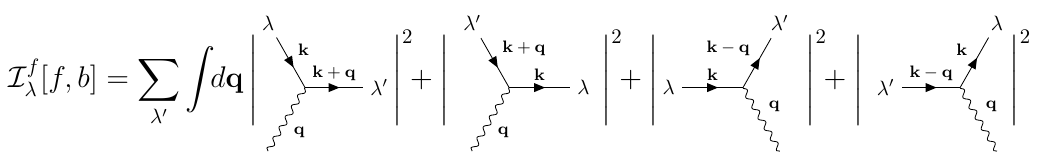}
    \includegraphics[scale=.5]{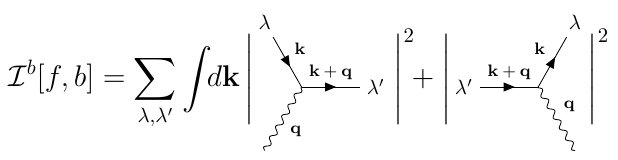}
    \caption{Diagrammatic representation of Eqs.\eqref{eq:colferm} and \eqref{eq:colbos}where each diagram reads as one of the term between square brackets.}
    \label{fig:my_label}
\end{figure}

\section{Joint fluid velocity from conservation laws}\label{app:htheorem}
The presence of a joint fluid velocity of fermions and plasmons assumed in the main text is traced back to the conservation of the total momentum. As the total momentum of the fluid is conserved, it gives rise to a single canonically conjugate hydrodynamic variable, namely the fluid velocity $\vec{u}$.
This can be proven by the means of entropy production, following Ref.\cite{Levchenko2020}.
The total entropy production $Q$ is given by

\begin{eqnarray}
    \label{EntProd}
    Q = \frac{dS}{dt} &=& \sum_\lambda\int d\vec{k}\log\left(\frac{1}{f_\lambda(\vec{k})}-1\right)\frac{\partial f_\lambda(\vec{k})}{\partial t}  \nonumber \\&&
    + \int d\vec{q}\log\left(\frac{1}{b(\vec{q})}+1\right)\frac{\partial b(\vec{q})}{\partial t} \geq 0, 
\end{eqnarray}
which cannot be negative according to the second law of thermodynamics and vanishes in thermal equilibrium.
The distribution functions, $f_\lambda(\vec{k})$ and $b(\vec{q})$, must obey the coupled Boltzmann equations in Eq.\eqref{eq:coupledboltzmann}with the scattering integrals given by Eqs. \eqref{eq:colbos},\eqref{eq:colferm}.
By making use of the Boltzmann equations, we can express the entropy production in Eq.(\ref{EntProd}) as:
\begin{multline}
    \label{EntProd2}
    Q = -\sum_\lambda\int d\vec{k}\log\left(\frac{1}{f_\lambda(\vec{k})}-1\right)\mathcal{I}^f_{\lambda}[f,b]  + \\
    - \int d\vec{q}\log\left(\frac{1}{b(\vec{q})}+1\right)\mathcal{I}^b[f,b]
\end{multline}

It is easy to see that the (local) equilibrium distribution functions that nullify the entropy production are given by
\begin{eqnarray}
    \label{DistFuncs}
    f^0_\lambda(\vec{k},\vec{r}) = \frac{1}{e^{\beta(\vec{r})(\epsilon_{\vec{k},\lambda}-\mu(\vec{r})-\vec{u}(\vec{r})\cdot\vec{k})}+1}, \\
    b^0(\vec{q},\vec{r}) = \frac{1}{e^{\beta(\vec{r})(\omega_\vec{q}-\vec{u}(\vec{r})\cdot\vec{q})}-1}.
\end{eqnarray}

The fluid velocity $\mathbf{u}(\mathbf{r})$ is the same for all three species.
It should be mentioned that the presence of the joint temperature $\beta(\vec{r})$ of fermions and plasmons is a consequence of total energy conservation.

\section{Plasmons and double-counting problem}
\label{app:double_count}
 At first glance, considering  electrons and plasmons as independent degrees of freedom seems like double-counting. It is true that we have derived the effective description of plasmon dynamics by integrating out the fermionic fields. However, based on the Keldysh QFT analysis, we have argued in Ref. \cite{Kitinan2022c} that this does not lead to double-counting. In fact, we have shown that considering both plasmons and electrons as independent entities is necessary for having a conservative approximation scheme \cite{KadanoffBaym1961,Narozhny2023}.

In this section, we provide an answer to this concern. We will give an additional justification for the coexistence of electrons and plasmons by calculating an equilibrium partition function of a system of Coulomb interacting electrons. We will show that the partition functions comprises two terms clearly identifiable as electron and plasmon contributions. 

To this end, we consider the equilibrium partition function reading as
\begin{equation}
Z = \int d[\psi^\dagger]d[\psi] e^{-S_0[\psi,\psi^\dagger]/\hbar-S_{\text{int}}[\psi,\psi^\dagger]/\hbar}.
\end{equation}
For simplicity, we consider an interacting 2DEG with a parabolic energy dispersion $\epsilon_{\vec{k}} = k^2/2m$. A generalization to graphene is straighforward. 
\begin{widetext}
Here
\begin{equation}
S_0[\psi,\psi^\dagger]= \int_0^{\hbar \beta} d\tau \int d\vec{x} ~ \psi(\vec{x},\tau) \underbrace{\left( -\hbar \partial_\tau - \frac{\hbar^2\vec{\nabla}^2}{2m}+V_{\text{ex}}(\vec{x})-\mu\right)}_{G^{-1}_0} \psi(\vec{x},\tau),
\end{equation}
and 
\begin{equation}
S_{\text{int}}[\psi,\psi^\dagger]= \frac{1}{2}\int_0^{\hbar \beta} d\tau \int d\vec{x} \int d\pvec{x}'~ \psi^\dagger(\vec{x},\tau)\psi(\vec{x},\tau) V(\vec{x}-\pvec{x}') \psi^\dagger(\pvec{x}',\tau)\psi(\pvec{x}',\tau).
\end{equation}
With the help of a Hubbard-Stratonovich transformation
\begin{equation}
e^{-S_{\text{int}}[\psi,\psi^\dagger]/\hbar}=\int d[\phi] e^{\frac{1}{2\hbar}\int d\tau \int d\vec{x}\int d \pvec{x}'\phi(\vec{x},\tau) V^{-1}(\vec{x}-\pvec{x}') \phi(\pvec{x}',\tau)  + \frac{1}{\hbar}\int d\tau d\vec{x} \phi(\vec{x},\tau) \psi(\vec{x},\tau) \psi^\dagger(\vec{x},\tau)},
 \end{equation}
 the fourth-order term in the fermionic field coming from the Coulomb interaction is traded for a quadratic theory of a real scalar field called plasmon field. At this point, the theory is rewritten exactly, no approximations are made. 
\end{widetext}

After the Hubbard-Stratonovich transformation, the theory becomes quadratic in fermionic fields. This allows us to integrate the fermions out exactly. This gives
\begin{equation}
Z = \int d[\phi] e^{S_{\text{eff}}},
\end{equation}
with
\begin{eqnarray}
S_{\text{eff}}&=&\frac{1}{2\hbar}\int d\tau \int d\vec{x}\int d \pvec{x}'\phi(\vec{x},\tau) V^{-1}(\vec{x}-\pvec{x}') \phi(\pvec{x}',\tau) \nonumber\\ &&+\Tr\left[\log\left(G^{-1}_0-\phi\right)\right].
\end{eqnarray}

To obtain the partition function, in principle, we have to proceed integrating out the plasmon field. This is however impossibly complicated, so an approximation is needed. To this end, we consider a fluctuation $\phi'$ of the real scalar field around its saddle point $\langle \phi \rangle$. We write $\phi = \langle \phi \rangle + \phi'$ and expand the effective action $S_{\text{eff}}$ up to second order in $\phi'$. We find that the saddle point solution $\langle\phi\rangle$ is determined from a self-consistent equation
$\langle \phi (\vec{x})\rangle = \int d\vec{x}' V(\vec{x}-\vec{x}') \int \frac{d\vec{p}}{(2\pi)^2} \sum_{\omega_n} \frac{1}{i\nu_n-\epsilon(\vec{p})+V_{\text{ex}}(\vec{x})+\langle \phi(\vec{x})\rangle}$ with $\nu_n$ being a fermionic Matsubara frequency. It is easy to see that $\langle \phi \rangle = -n_0\int d\vec{x}' V(\vec{x}-\vec{x}')$.  This implies that  $\langle  \phi \rangle$ is canceled exactly by $V_{\text{ex}}(\vec{x})$ coming from the positive charge background.
\begin{widetext}
After expanding the effective action up to quadratic order, we find that 
\begin{equation}
S_{\text{eff}} \approx \frac{1}{2\hbar}\int d\tau \int d\vec{x}\int d \pvec{x}'\phi'(\vec{x},\tau) V^{-1}(\vec{x}-\pvec{x}') \phi'(\pvec{x}',\tau) +\Tr\left[\log\left(-G^{-1}_{\langle\phi\rangle}\right)\right]-\frac{1}{2}\Tr\left[G_{\langle\phi\rangle}\phi' G_{\langle\phi\rangle}\phi' \right].
\end{equation}
\end{widetext}
Here
\begin{equation}
G^{-1}_{\langle  \phi\rangle}(\vec{x},t;\pvec{x}',t') = \delta(\vec{x}-\pvec{x}')\delta(t-t')\left( -\hbar \partial_\tau - \frac{\hbar^2\vec{\nabla}^2}{2m}-\mu\right).
\end{equation}
As the theory becomes quadratic in the plasmon field $\phi'$, it allows us to integrate over this field to obtain an approximate partition partition function reading as
\begin{eqnarray}
\label{eq:partapprox}
Z &\approx&  \exp\left(-\frac{1}{2} \sum_{m} \int_{\vec{q}} \log\left(-V^{-1}(\vec{q})+\Pi(i\omega_m,\vec{q})\right) \right) \nonumber\\ && \times \exp\left(\sum_n \int_{\vec{k}}\log\left(G^{-1}_{\langle\phi\rangle}\left(i\nu_n,\vec{k}\right)\right) \right),
\end{eqnarray}
with $\Pi(i\omega_m,\vec{q}) = \frac{1}{\beta}\sum_{n} \int_{\vec{k}}G_{\langle\phi\rangle}(\vec{k}+\vec{q},i\omega_m+i\nu_n)G_{\langle \phi \rangle}(\vec{k},i\nu_n) $  defining the polarization function. For 2DEGs, the polarization function in the long-wavelength limit is given by 
$\Pi(i\omega_n,\vec{q}) \approx \frac{q^2\mu}{2\pi(i\omega_n)^2}$  \cite{Stern1967}. Here $\omega_n$ denotes the bosonic Matsubara frequency.

The summations over Matsubara frequencies in Eq. \ref{eq:partapprox} can be evaluated using a standard contour integration technique. We refer the readers to, e.g, Ref. \cite{Stoof2009} for details. After straightforward steps, we find that
\begin{eqnarray}
Z &=& \exp\left(-\int_{\vec{q}}\log(1-e^{-\beta \omega_{\text{pl}}(\vec{q})})\right) \nonumber\\ &&\times \exp\left(\int_{\vec{p}}\log(1+e^{-\beta(\epsilon(\vec{p})-\mu)})\right).
\end{eqnarray}
This partition function describes a combined system of bosons of energy $\omega_{\text{pl}}(\vec{q})$ and fermions of energy $\epsilon(\vec{p})=p^2/2m$. The energy dispersion $\omega_{\text{pl}}(\vec{q})$ of plasmons is determined from $-V^{-1}+\Pi=0$. Within this approximation the system consists of two independent degrees of freedom, i.e., electrons and plasmons.  Those are proper quasiparticles in their own right.

\end{document}